\documentclass[aps, showpacs]{revtex4}
\usepackage{graphicx}
\begin{document}

\title{Oblique amplitude modulation of dust-acoustic plasma waves
\footnote{Preprint, submitted to Physica Scripta.}}

\author{I. Kourakis}
\altaffiliation[On leave from: ]{U.L.B. - Universit\'e Libre de
Bruxelles, Facult\'e des Sciences Apliqu\'ees - C.P. 165/81
Physique G\'en\'erale, Avenue F. D. Roosevelt 49, B-1050 Brussels,
Belgium}
\email{ioannis@tp4.rub.de}
\author{P. K. Shukla}%
\affiliation{Institut f\"ur Theoretische Physik IV, Fakult\"at f\"ur Physik
und Astronomie, Ruhr--Universit\"at Bochum, D--44780 Bochum, Germany}

\date{\today}


\begin{abstract}
Theoretical and numerical studies are presented of the nonlinear
amplitude modulation of dust-acoustic (DA) waves propagating in an
unmagnetized three component, weakly-coupled, fully ionized plasma
consisting of electrons, positive ions and charged dust particles,
considering perturbations oblique to the carrier wave propagation
direction. The stability analysis, based on a nonlinear
Schr\"odinger-type equation (NLSE), shows that the wave may become
unstable; the stability criteria
 depend on the angle $\theta$ between the
modulation and propagation directions. Explicit expressions for
the instability rate and threshold have been obtained in terms of the
dispersion laws of the system.
The possibility and conditions for the existence of different types of
localized excitations have also been discussed.
\end{abstract}

\pacs{52.27.Lw, 52.35.Fp, 52.35.Mw, 52.35.Sb}

\maketitle


\section{Introduction}

The study of the dynamics of dust contaminated plasmas (DP) has
recently received considerable interest due to their occurrence in
real charged particle systems, e.g. in space and laboratory
plasmas and the novel physics involved in their description
\cite{PSbook}. An issue of particular interest is the existence of
special acoustic-like oscillatory modes, e.g. the dust-acoustic
waves (DAW) and dust-ion-acoustic waves (DIAW), which were
theoretically predicted about a decade ago \cite{Rao, SSIAW} and
later experimentally confirmed \cite{Barkan, Pieper}. The DAW,
which we consider herein, relies on a new physical mechanism in
which inertial dust grains oscillate against a thermalized
background of electrons and ions which provide the necessary
restoring force. The phase speed of the DAW is much smaller than
the electron and ion thermal speeds, and the DAW frequency is
below the dust plasma frequency.

A long-known generic characteristic of nonlinear wave propagation
is amplitude modulation due to the nonlinear self-interaction of
the carrier wave, which generates higher harmonics. The standard
method for studying this mechanism is a multiple space and time
scale technique \cite{redpert, redpert2}, which leads to a
nonlinear Schr\"odinger-type equation (NLSE) describing the
evolution of the wave envelope. It has been shown that, under
certain conditions, waves may develop a Benjamin-Feir-type
(modulational) instability (MI), i.e. their modulated envelope may
collapse under the influence of external perturbations.
Furthermore, the NLSE-based analysis, already present in a wide
variety of contexts \cite{Remoissenet, Sulem, Hasegawa1}, reveals
the possibility of the existence of localized excitations
(solitary wave structures) whose form and behaviour depends on
criteria similar to the ones necessary for the MI to occur.

Not surprisingly, plasma wave theory has provided an excellent
test bed for this approach since a long time ago [7, 11 - 19]
and dusty plasma waves were no exception [20 -- 22].
Among other noteworthy results, electron plasma modes have been
shown to be stable against parallel modulation \cite{Kakutani}; so
do the ion plasma modes, yet only for perturbations below a
specific wavenumber threshold \cite{Chan}. Electron and ion
acoustic modes, even though stable to parallel modulation
\cite{Kakutani, Shimizu, Kako1, comment1}, are found to be
unstable if one takes into account finite temperature effects
\cite{Chan, Durrani, chin3} or, most interesting to us, when
subject to an oblique modulation of the wave amplitude [17 -- 19].
These results, based on Poisson - moment plasma equations, have
been confirmed by similar studies from a kinetic point of view
\cite{kinetic}, for the ion - acoustic wave in an electron - ion
plasma. In dusty plasma, the amplitude modulation of the DAW and
DIAW has been investigated in Ref. [20 -- 22];
similar studies have been carried out for oscillations in
(strongly-coupled) dusty plasma quasi-crystals [25 - 26].
Finally, let us mention that attempts have
been made to refine the description of the DIAW modulation by
including non-planar geometry effects \cite{chin2}, following an
idea applied earlier in the KdV (Korteweg-de Vries) description of
a dusty plasma \cite{MS},
 and dust-charge fluctuation effects \cite{chin},
 an issue of particular importance in the present-time DP surveys
(see e.g. \cite{Ivlev, Mamun}; also \cite{PSbook}). These effects,
omitted in the present investigation, will be considered in a
forthcoming work.


In this paper, we study the modulational instability of dust-acoustic
plasma waves propagating in an unmagnetized plasma contaminated by
a population of charged dust grains, whose dimensions and charge are
assumed constant, for simplicity.  Amplitude modulation is allowed to
take place in an oblique direction, at an angle $\theta$ with respect
to the carrier wave propagation direction.  Once an explicit criterion for the
occurrence of instability is established, our aim is to trace the influence of
$\theta$ on the conditions for the MI onset, and determine the magnitude
of the associated instability growth rate. Finally, we shall also examine the
possibility of the formation of localized excitations and discuss their
characteristics.  Exact new expressions are derived for all quantities of interest,
in terms of the system's dispersion laws.  Among other physical parameters discussed,
our formulation leaves open the choice of {\em sign} of dust charge ($sign \,q_d = \pm 1$)
(most often taken to be negative since this is the most frequently occurring case
\cite{PSbook}) and the dust pressure (`temperature') scaling.
Our aim in doing so is to address, among others, the question of the
influence of the dust charge {\em sign} on the amplitude modulation mechanism.
We may also attempt to clarify the effect of taking (or not) into account
the dust pressure evolution equation (omitted e.g. in \cite{AMS}) in the analysis.

The manuscript is organized as follows. In the next Section,
the analytical model is introduced. In Section III, we carry out a
perturbative analysis by introducing appropriate slow space and time evolution
scales, and derive a NLS-type equation which governs the (slow)
amplitude evolution in time and space. The exact form of dispersion
and nonlinearity coefficients in the NLS-type equation is presented and
discussed. In Section IV, we carry out a stability
analysis of the NLSE allowing for a thorough study of the DAW
stability in various regions of the physical parameters involved.
The analysis is pursued in Section V, where we discuss the
possibility of the existence of localized solutions of the NLSE, and
identify their forms in different parameter regions. Finally, we
briefly summarize our results in the concluding Section.

\section{The model}

We consider a three component collisionless unmagnetized dusty
plasma consisting of electrons (mass $m$, charge $e$), ions
(mass $m_i$, charge $q_i = + Z_i e$) and heavy dust particulates
(mass $m_d$, charge $q_d = s\, Z_d e$), henceforth denoted by
$e,\, i, \,d$ respectively. Dust mass
and charge will be taken to be constant, for simplicity. Note that
both negative and positive dust charge cases are considered,
distinguished by the charge sign $s = sgn \,q_d = \pm 1$ in the
formulae below.

\subsection{Evolution equations}

The basis of our study includes the moment - Poisson system of equations for
the dust particles and Boltzmann distributed electrons and ions.
The dust (number) density $n_d$ is governed by the (continuity)
equation
\begin{equation}
\frac{\partial n_d}{dt} + \nabla \cdot (n_d \,\mathbf{u}_d)= 0 \, ,
\label{densityequation}
\end{equation}
and the dust mean velocity $\mathbf{u}_d$ obeys
\begin{equation}
\frac{\partial \mathbf{u}_d}{dt} + \mathbf{u}_d \cdot \nabla
\mathbf{u}_d \, = \, - \frac{q_d}{m_d}\,\nabla \,\Phi -
\frac{1}{m_d n_d}\,\nabla p_d \, ,
\end{equation}
where $\Phi$ is the electric potential. The dust pressure $p_d$
obeys
\begin{equation}
\frac{\partial p_d}{dt} + \mathbf{u}_d \cdot \nabla p_d \, = \, -
\gamma\, p_d \,\nabla \cdot \mathbf{u}_d \label{eqp} \,
\end{equation}
Here $\gamma = (f + 2)/f$ is the ratio of specific heats ($f$ is the number of
degrees of freedom) e.g. $\gamma = 3$ in the adiabatic one-dimensional (1d) case
and $\gamma = 2$ in the two-dimensional (2d) case.
The system is
closed with Poisson's equation
\begin{eqnarray}
\nabla^2 \Phi \, =\, - 4 \pi \, \sum q_s\, n_s \, &= & \,4 \pi
\,(n_e \,e - q_i \,n_i - q_d\,n_d)
\equiv \,4 \pi \,e \,(n_e  - Z_i \,n_i - s \,Z_d\,n_d)  \, ;
\label{Poisson}
\end{eqnarray}
note that the right-hand-side cancels at equilibrium due
to the overall neutrality condition
\begin{equation}
n_{e,0} \,e - n_{i, 0} \,q_i - n_{d, 0} \,q_d \, =\, 0 \, .
\label{neutrality}
\end{equation}
The right-hand side in (\ref{Poisson}) is often formulated in terms of the
ratio $\mu = n_{e,0}/(Z_i n_{i,0})$; for convenience, we have
\begin{equation} \mu \, = \, 1 + s\, \frac{Z_{d}}{Z_{i}}
\frac{n_{d,0}}{n_{i,0}}
\end{equation}
due to (\ref{neutrality}), so that a value lower (higher) than $1$
corresponds to negative (positive) dust charge; $\mu$ obviously
tends to unity in the absence of dust (in any case, $\mu > 0$). We
will retain this notation in the following, for the sake of
reference to previous works.

The electrons and ions are assumed to be close to a Maxwellian
equilibrium. The corresponding densities are
\[n_e \approx n_{e,0}\, e^{e \Phi/k_B T_e} \, ,\]
and
\begin{equation}
n_i \approx  n_{i,0}\, e^{- Z_i\, e \Phi/k_B T_i} \, ,
\label{Maxwell}
\end{equation}
where $T_s$ denotes the
temperature of species $s = e, i$ ($k_B$ is the Boltzmann constant).

\subsection{Reduced equations}

Re-scaling all variables over appropriately chosen quantities and
developing around $\Phi = 0$, Eqs. (\ref{densityequation}) - (\ref{Maxwell}) can be
cast in the reduced form
\begin{eqnarray}
\frac{\partial n}{dt} + \nabla \cdot (n \,\mathbf{u})&= & 0\, ,
\nonumber \\
\frac{\partial \mathbf{u}}{dt} + \mathbf{u}  \cdot \nabla \mathbf{u} \,
&=&
\, - s\,\nabla \phi - \frac{\sigma}{n}\,\nabla p \, , \nonumber \\
\frac{\partial p}{dt} + \mathbf{u}  \cdot \nabla p \, &= &\, - \gamma\,
p
\,\nabla  \cdot \mathbf{u} \, ,\nonumber
\end{eqnarray}
and
\begin{equation}
\nabla^2 \phi \, =\, \phi - \alpha \,\phi^2 + \alpha' \,\phi^3 -
s \,\beta\,(n - 1)\, ,
\label{reducedeqs}
\end{equation}
where all quantities are non-dimensional: $n = n_d/n_{d, 0}$,
$\mathbf{u} = \mathbf{u}_d/v_{0}$, $p = p_d/p_{0}$ and
$\phi = \Phi/\Phi_0$; the scaling quantities (index $0$)
are, respectively: the equilibrium density $n_{d, 0}$,
the `dust sound speed' $v_{0} = v_d = (k_B T_e/m_d)^{1/2}$,
$p_0 = n_{d, 0} k_B T_e$ and
$\Phi_{0} = (k_B T_e/Z_d e)$.
Space and time in (\ref{reducedeqs}) are, respectively, scaled
over: the DP effective Debye length
$\lambda_{D, eff} = (\lambda_{D, e}^{-2}+\lambda_{D, i}^{-2})^{-1/2}$
(where $\lambda_{D, s} = (k_B T_s/4 \pi n_{s, 0} q_{s}^2)^{1/2}$, $s = e, i$)
and the inverse DP plasma frequency
$\omega_{p, d}^{-1} = (4 \pi n_{d, 0} q_{d}^2/m_d)^{- 1/2}$.
Recall that $s = $sgn $q_d$, so the influence of the dust charge
{\em {sign}} will be traced via the appearance of $s$ in the
forthcoming formulae. Finally, $\sigma = p_0/(n_{d, 0} k_B T_e)$
is equal to unity, given the above choice for $p_0$; nevertheless,
$\sigma$ - often interpreted as a temperature ratio via a
different scaling, see e.g. \cite{chin} - will be retained in
order to `tag' the influence of the coupling to pressure evolution
equation (\ref{eqp}) being taken into account - as compared to
a previous work \cite{AMS} where Eq. (\ref{eqp})
has been omitted. As a matter of fact, expressions (9) - (11) therein are readily
recovered here upon setting $\sigma = 0$, $s = -1$, $\alpha' = 0$ in
Eq. (\ref{reducedeqs}) above.

The dimensionless parameters appearing in (\ref{reducedeqs}) are
\[
\alpha = \frac{1}{2 Z_d}
\frac{Z_i^3 (\frac{T_e}{T_i})^2 \frac{n_{i, 0}}{n_{e, 0}} -1}
{Z_i^2 \frac{T_e}{T_i}
\frac{n_{i, 0}}{n_{e, 0}} + 1}
\, , \qquad
\alpha' = \frac{1}{6 Z_d^2}
\frac{Z_i^4 (\frac{T_e}{T_i})^3 \frac{n_{i, 0}}{n_{e, 0}} + 1}
{Z_i^2 \frac{T_e}{T_i}
\frac{n_{i, 0}}{n_{e, 0}} + 1} \, ,
\]
and
\[
\beta = \biggl( \lambda_{D, eff} \frac{\omega_{p, d}}{v_d} \biggl)^2
\equiv
\biggl( \frac{c_{D}}{v_d} \biggl)^2 \, ,
\]
where $c_{D} = {\lambda_{D}}_{eff} \omega_{p, d}$ is the DA speed
\cite{PSbook}. Alternatively, in terms of $\mu$ defined above, one
has: \( \alpha = - \frac{1}{2} \frac{\mu \alpha_1^2 -
\alpha_2^2}{\mu \alpha_1 + \alpha_2} \), \( \alpha' = \frac{1}{6}
\frac{\mu \alpha_1^3 + \alpha_2^3}{\mu \alpha_1 + \alpha_2} \), \(
\beta = s\, \frac{\mu - 1}{\mu \alpha_1 + \alpha_2} \), where \(
\alpha_1 = \frac{1}{Z_d} \) and \( \alpha_2 = \frac{Z_i}{Z_d}
\frac{T_e}{T_i} \). All these parameters are positive
\cite{comment2}. For $\mu \ll \frac{\alpha_2}{\alpha_1} = Z_i
\,\frac{T_e}{T_i}$, we have the approximate expressions: \( \alpha
\approx \frac{\alpha_2}{2} = \frac{Z_i}{2 Z_d} \frac{T_e}{T_i} \)
and \( \alpha' \approx \frac{\alpha_2^2}{6} = \frac{Z_i^2}{6
Z_d^2} \frac{T_e^2}{T_i^2} \approx \, \frac{2}{3} \, \alpha^2 \),
as in \cite{AMS}; \, also: $\beta \approx
\frac{Z_d^2}{Z_i^2}\frac{n_{d, 0}}{n_{i, 0}} \frac{T_i}{T_e}$. A
comment should be made, regarding the order of magnitude of the
parameters $\alpha$, $\alpha'$, $\beta$. Notice that $\alpha$
takes very small (positive) values (as low as, say, $10^{-4}$ to
$10^{-2}$) and so does $\alpha'$; however, $\beta$ may take high
values, e.g. ranging from zero (for $\mu = 1$ i.e. no dust) to,
say, $10^{2} - 10^{3}$. Therefore, the numerical result of the
scaling in our (DAW) case is completely different from the one in
the dust ion-acoustic (DIAW) case \cite{IKPSDIAW}, despite the
apparent similarity in the model expressions \cite{AMS,
commentdiaw}; this is why we chose not to analyse the DIAW case
any further, in the same text.

\section{Perturbative analysis}

\subsection{Outline of the method}

Let $S$ be the state (column) vector $(n, \, \mathbf{u} , \, p, \, \phi)^T$,
describing the system's state at a given position $\mathbf{r}$ and instant $t$.
We shall consider small deviations from the equilibrium state
$S^{(0)} = (1, \, \mathbf{0} , \, 1, \,0)^T$ by taking
\[
S = S^{(0)} \, + \, \epsilon \, S^{(1)} + \, \epsilon^2 \, S^{(2)} + \, ... =
S^{(0)} \, + \,
\, \sum_{n=1}^\infty \epsilon^n \, S^{(n)} \, ,
\]
where $\epsilon \ll 1$ is a smallness parameter.
Following the standard multiple scale (reductive perturbation)
technique \cite{redpert}, we shall
consider the following stretched (slow) space and time variables
\begin{equation}
\zeta \,= \, \epsilon (x - \lambda \,t) \, , \qquad
\tau \,= \, \epsilon^2 \, t\, ,
\label{slowvar}
\end{equation}
where $\lambda$, bearing dimensions of velocity,
is to be later interpreted as the group velocity in the $x$ direction.
In order to take into account the influence of an oblique amplitude modulation
on the DA
wave, we will assume that all perturbed states depend on the fast scales
via the phase $\theta_1 = \mathbf{k \cdot r} - \omega t$ only, while the slow
scales enter the argument
of the $l-$th harmonic amplitude $S_l^{(n)}$, which is allowed to vary along
$x$,
\[
S^{(n)} \,= \,
\sum_{l=-\infty}^\infty \,S_l^{(n)}(\zeta, \, \tau)
 \, e^{i l (\mathbf{k \cdot r} - \omega t)} \, .
\]
The reality condition $S_{-l}^{(n)} = {S_l^{(n)}}^*$ is met by
all state variables.
Note that the (choice of) direction of the propagation remains
arbitrary,
yet modulation is allowed to take place in an oblique direction,
characterized by a
pitch angle $\theta$. Assuming the modulation direction to define the $x-$ axis,
the wave-number vector
$\mathbf{k}$ is taken to be
\(
\mathbf{k} = (k_x, \, k_y) = (k\, \cos\theta, \, k\, \sin\theta)
\). According to these considerations,
the derivative operators in the above equations are treated as follows
\[
\frac{\partial}{\partial t} \rightarrow \frac{\partial}{\partial t} -
\epsilon \, \lambda \, \frac{\partial}{\partial \zeta} +
\epsilon^2 \, \frac{\partial}{\partial \tau} \, ,
\]
\[
\nabla \rightarrow \nabla +
\epsilon \, \hat x \, \frac{\partial}{\partial \zeta} \, ,
\]
and
\[
\nabla^2 \rightarrow \nabla^2 +
2 \epsilon \, \frac{\partial^2}{\partial x \partial\zeta} +
\epsilon^2 \, \frac{\partial^2}{\partial\zeta^2} \, ,
\]
i.e. explicitly
\[
\frac{\partial }{\partial t} \, A_l^{(n)}
\, e^{i l \theta_1}
= \biggr( - i l \omega \, A_l^{(n)}
\, -
\epsilon \, \lambda \, \frac{\partial A_l^{(n)}}{\partial \zeta} +
\epsilon^2 \, \frac{\partial A_l^{(n)} }{\partial \tau} \biggr)
\, e^{i l \theta_1} \, ,
\]
\[
\nabla \, A_l^{(n)}
\, e^{i l \theta_1}
= \biggr( i l \mathbf{k} \, A_l^{(n)}
\, +
\epsilon \, \hat x \, \frac{\partial A_l^{(n)}}{\partial \zeta}  \biggr)
\, e^{i l \theta_1}  \, ,
\]
and
\[
\nabla^2 A_l^{(n)}\, e^{i l \theta_1}
= \biggr( - l^2 k^2 \, A_l^{(n)}
\, +
2 \epsilon \, i l k_x
\, \frac{\partial A_l^{(n)}}{\partial\zeta} +
\epsilon^2 \, \frac{\partial^2 A_l^{(n)}}{\partial\zeta^2}
\biggr)
\, e^{i l \theta_1}
\]
for any $A_l^{(n)}$ of the components of $S_l^{(n)}$.

\subsection{Amplitude evolution equations}

By substituting the above expressions into the system of equations
(\ref{reducedeqs}) and isolating distinct orders in $\epsilon$, we obtain
the $n$th-order reduced equations
\begin{eqnarray}
- i l \omega n_l^{(n)} \,+\, i l \mathbf{k \cdot u}_l^{(n)} \,-\,
\lambda \, \frac{\partial n_l^{(n-1)}}{\partial \zeta}
\,+\,
\frac{\partial n_l^{(n-2)}}{\partial \tau}
\,+\,
\frac{\partial u_{l, x}^{(n-1)}}{\partial \zeta} \qquad \qquad \qquad
\qquad \qquad \qquad
\nonumber \\
\,+\, \sum_{n' = 1}^{\infty} \, \sum_{l' = -\infty}^{\infty}
\biggl[  i l \mathbf{k \cdot u}_{l-l'}^{(n-n')} \, n_{l'}^{(n')} +\,
\frac{\partial}{\partial \zeta}
\biggl( n_{l'}^{(n')} u_{(l-l'), x}^{(n-n'-1)}\biggr)
\biggr]  \, = \, 0 \, ,\quad
\label{geneqn1} \\
\nonumber \\
- i l \omega \mathbf{u}_l^{(n)} \,+\, s \, i l \mathbf{k} \phi_l^{(n)} \,-\,
\lambda \, \frac{\partial \mathbf{u}_l^{(n-1)}}{\partial \zeta}
\,+\,
\frac{\partial \mathbf{u}_l^{(n-2)}}{\partial \tau}
\,+s\,
\frac{\partial \phi_{l}^{(n-1)}}{\partial \zeta} \,\hat x
\qquad \qquad \qquad  \qquad \qquad \qquad
\nonumber \\
\,+\, \sum_{n' = 1}^{\infty} \, \sum_{l' = -\infty}^{\infty}
\biggl[  i l' \mathbf{k \cdot u}_{l-l'}^{(n-n')} \, \mathbf{u}_{l'}^{(n')} +\,
 u_{(l-l'), x}^{(n-n'-1)}\,
\frac{\partial \mathbf{u}_{l'}^{(n')}}{\partial \zeta}
\biggr]
\nonumber \\
+ \, \sigma \, \biggl( il p_{l}^{(n)}\, \mathbf{k}\,+\,
\frac{\partial p_l^{(n-1)}}{\partial \zeta} \, \hat x \biggr)
\nonumber \\
+ \,
\sum_{n' = 1}^{\infty} \, \sum_{l' = -\infty}^{\infty}
n_{(l-l')}^{(n-n')} \,
\biggl\{
- i l' \omega \mathbf{u}_{l'}^{(n')} \,
+\, s \, i {l'} \mathbf{k} \phi_{l'}^{(n')} \,
-\,
\lambda \, \frac{\partial \mathbf{u}_{l'}^{(n'-1)}}{\partial \zeta}
\,+\,
\frac{\partial \mathbf{u}_{l'}^{(n'-2)}}{\partial \tau}
\,
+s\,
\frac{\partial \phi_{l'}^{(n'-1)}}{\partial \zeta} \,\hat x
\nonumber \\
+ \,
\sum_{n'' = 1}^{\infty} \, \sum_{l'' = -\infty}^{\infty}
\biggl[  i l'' \mathbf{k \cdot u}_{l'-l''}^{(n'-n'')} \,
\mathbf{u}_{l''}^{(n'')} +\,
 u_{(l'-l''), x}^{(n'-n''-1)}\,
\frac{\partial \mathbf{u}_{l''}^{(n'')}}{\partial \zeta}
\biggr]
\biggr\} \, = \, 0 \, ,\quad
\label{geneqn2} \\
\nonumber \\
- i l \omega p_l^{(n)} \,+\,  i l \gamma \, \mathbf{k\cdot u}_l^{(n)} \,-\,
\lambda \, \frac{\partial p_l^{(n-1)}}{\partial \zeta}
\,+\,
\frac{\partial p_l^{(n-2)}}{\partial \tau}
\,+\,
\gamma \, \frac{\partial u_{l, x}^{(n-1)}}{\partial \zeta}
\qquad \qquad \qquad \qquad \qquad \qquad
\nonumber \\
\,+\, \gamma \,
\sum_{n' = 1}^{\infty} \, \sum_{l' = -\infty}^{\infty}
p_{l-l'}^{(n-n')} \,
\biggl(  i l' \mathbf{k \cdot u}_{l'}^{(n')} +\,
\frac{\partial u_{l', x}^{(n'-1)}}{\partial \zeta}
\biggr)
\nonumber \\
\,+\, \sum_{n' = 1}^{\infty} \, \sum_{l' = -\infty}^{\infty}
\,
\biggl(  i l' \mathbf{k \cdot u}_{l-l'}^{(n-n')} p_{l'}^{(n')} +\,
\frac{\partial u_{l'}^{(n'-1)}}{\partial \zeta}  u_{(l-l'), x}^{(n-n')}
\biggr) \, = \, 0 \, ,\quad \label{geneqn3}
\end{eqnarray}
and
\begin{eqnarray}
- (l^2 k^2 + 1)\, \phi_l^{(n)} \,+ \,s\,\beta \, n_l^{(n)}
+\,  2 i l k_x \, \frac{\partial \phi_l^{(n-1)}}{\partial \zeta}
\,+\,
\frac{\partial^2 \phi_l^{(n-2)}}{\partial \zeta^2}
\qquad \qquad \qquad \qquad \qquad \qquad
\nonumber \\
+\, \alpha \,\sum_{n' = 1}^{\infty} \, \sum_{l' = -\infty}^{\infty}
\,
\phi_{l-l'}^{(n-n')} \,\phi_{l'}^{(n')} \,
- \, \alpha' \,\sum_{n', n'' = 1}^{\infty} \,
\sum_{l', l'' = -\infty}^{\infty}
\,
\phi_{l-l'-l''}^{(n-n'-n'')} \,\phi_{l'}^{(n')}\,\phi_{l''}^{(n'')}
 \, = \, 0 \, . \quad
\label{geneqn4}
\end{eqnarray}
Notice the last three lines in Eq. (\ref{geneqn2}), which are due to the
consideration of the pressure evolution equation (\ref{eqp}) and are
absent e.g. in Ref. \cite{AMS} - cf. Eq. (33) therein.  Even though it is
$\sigma$ which introduces coupling to (\ref{geneqn3}) (which
becomes decoupled from the rest, that is, for $\sigma = 0$), should one
correctly consider the limit $\sigma = 0$ in Eqs.
(\ref{reducedeqs}), one should discard all three of the
last lines in (\ref{geneqn2}). For convenience, one may consider
instead of the vectorial relation (\ref{geneqn2}) the one obtained
by taking its scalar product with the wavenumber $\mathbf{k}$.
Finally, we see that Eqs. (32) - (34) of Ref. \cite{AMS} are readily recovered
upon setting $\sigma = 0$, $s = -1$ and $\alpha' = 0$ in the above
relations.

\subsection{First order in $\epsilon$:
first harmonics and dispersion relation}

The first order ($n = 2$) equations read
\begin{eqnarray}
- i l \omega n_l^{(1)} \,+\, i l \mathbf{k \cdot u}_l^{(1)} \, = \, 0 \, ,
\label{1eqn1}
\\
- i l \omega \mathbf{u}_l^{(1)} \,+\, s \, i l \mathbf{k} \phi_l^{(1)} \,
+ \, il \sigma \,  p_{l}^{(1)}\, \mathbf{k}\,= \, 0 \, ,
\label{1eqn2} \\
- i l \omega p_l^{(1)} \,
+\,  i l \gamma \, \mathbf{k\cdot u}_l^{(1)} \, = \, 0 \, ,
\label{1eqn3} \end{eqnarray}
and
\begin{equation}
- (l^2 k^2 + 1)\, \phi_l^{(1)} \,+ \,s\,\beta \, n_l^{(1)}
\, = \, 0 \, .
\label{1eqn4}
\end{equation}
For $l = 1$, these equations determine
the first harmonics of the perturbation.
The following dispersion relation is obtained
\begin{equation}
\omega^2\,  = \frac{\beta \, k^2}{k^2 + 1} \, +
\, \gamma \, \sigma \, k^2 \, .
\label{dispersion}
\end{equation}
Restoring dimensions, one may easily check that the standard DAW
dispersion relation \cite{PSbook, Rao} is thus exactly recovered:
\begin{eqnarray}
\omega^2\,  & = & \omega_{p, d}^2\, \frac{k^2}{k^2 + k_D^2 } \, +
\, \gamma \, \frac{k_B T_d}{m_d} \, k^2
\equiv \,\frac{c_{D}^2\, k^2}{1 + k^2 \, {\lambda_{D}}_{eff}^2} \,
+ \, \gamma \, v_{th, d}^2 \, k^2 \, . \label{dispersion-dim}
\end{eqnarray}

The first harmonic amplitudes may now be expressed in terms of the
first order potential correction $\phi_1^{(1)}$; we obtain the
relations
\begin{eqnarray}
n_1^{(1)} \,  & = & s\, \frac{1 + k^2}{\beta} \phi_1^{(1)} \equiv
c^{(11)}_1 \,\phi_1^{(1)} \, ,\nonumber \\
\mathbf{k\cdot u}_1^{(1)}\, &= &\,
\omega \, n_1^{(1)}
\, = s\,
\omega \frac{1 + k^2}{\beta} \, \phi_1^{(1)} \equiv c^{(11)}_2 \,\phi_1^{(1)}
\, ,
\nonumber \\
p_1^{(1)} \, &= & \gamma \, n_1^{(1)}
\, = \gamma \, s\,
\frac{1 + k^2}{\beta}  \phi_1^{(1)}  \equiv c^{(11)}_3
\,\phi_1^{(1)} \, , \, \nonumber \\
\qquad u_{1, x}^{(1)} \, & = &
\frac{\omega}{k} \cos\theta \,  n_1^{(1)}
\, = s\, \frac{1 + k^2}{\beta}
\frac{\omega}{k} \cos\theta \,  \phi_1^{(1)} = c^{(11)}_5
\,\phi_1^{(1)} \, , \qquad \nonumber \\
\qquad u_{1, y}^{(1)}  \, &= &
\frac{\omega}{k} \sin\theta \,  n_1^{(1)} \, = s\, \frac{1 +
k^2}{\beta} \frac{\omega}{k} \sin\theta \, \phi_1^{(1)} \, , \qquad
\label{coeffs11}
\end{eqnarray}
retaining, for later use, the (obvious) definitions of the coefficients
$c^{(11)}_j$ ($j = 1, ..., 5$) relating the state variables to the 1st-order
potential
correction $\phi_1^{(1)}$ (so $c_4^{(11)} = 1$).

\subsection{Second order in $\epsilon$:
group velocity, 0th and 2nd harmonics}

The second order ($n = 2$) equations for the first harmonics
provide the compatibility condition: \( \lambda \,  = v_g(k) \,  =
\frac{\partial \omega}{\partial k_x} = \omega'(k) \cos\theta =$ $
\frac{k}{\omega} \bigl[ \frac{1}{(1 + k^2)^2} + \gamma \sigma \bigr]
\cos\theta \); the
group velocity $v_g$ can be cast in the form
\begin{equation}
v_g(k) \,  =\, \frac{\omega^3}{k^3} \, \, \frac{\beta + \sigma
\gamma (1 + k^2)^2}{[\beta + \sigma \gamma (1 + k^2)]^2 } \,
\cos\theta\equiv  \, \frac{\omega^3}{\beta \, k^3}\,\nu_1\,
\cos\theta \, ,\label{vg}
\end{equation}
where we have denoted
\begin{equation}
\nu_1 = \beta \,\frac{\beta + \sigma \gamma (1 + k^2)^2}{[\beta +
\sigma \gamma (1 + k^2)]^2 } \, .
\label{defnu1}
\end{equation}
Note that $\nu_1 \rightarrow 1$ in the limit
$\sigma \rightarrow 0$, recovering exactly Eq. (43) in Ref. \cite{AMS}.

The 2nd-order corrections to the first
harmonic amplitudes are now given by
\begin{eqnarray}
n_1^{(2)} \,  & = & i\, s\, \frac{1}{\beta} \,\bigl[ \tilde A (1 + k^2) - 2
k \cos\theta \bigr] \,\frac{\partial \phi_1^{(1)}}{\partial \zeta}
\equiv i \, c^{(21)}_1 \,\frac{\partial \phi_1^{(1)}}{\partial
\zeta} \, ,\nonumber \\
\mathbf{k\cdot u}_1^{(2)}\, &=& \omega n_1^{(2)} - s  \frac{1}{\beta} \,
(1 + k^2) \,
 \biggl(v_g - \frac{\omega}{k}\cos\theta \biggr)
 \frac{\partial \phi_1^{(1)}}{\partial
\zeta}
 \equiv i \, c^{(21)}_2 \,\frac{\partial \phi_1^{(1)}}{\partial
\zeta} \, ,\nonumber \\
p_1^{(2)} \, &=& \gamma \, n_1^{(2)}\,\equiv \, i
\, c^{(21)}_3 \,\frac{\partial \phi_1^{(1)}}{\partial
\zeta} \, ,\nonumber
\\
 \qquad \phi_1^{(2)} &=& i \, \tilde A \,\frac{\partial
\phi_1^{(1)}}{\partial \zeta} \, ,\nonumber
\end{eqnarray}
and
\begin{eqnarray}
u_{1, x}^{(2)}\, &=& i\,s \frac{1}{\omega}\biggl[-1 - 2\,
\frac{\gamma}{\beta}\, \sigma \, k^2 \, \cos^2\theta \,+
\biggl(v_g\,  \frac{\omega}{k}\cos\theta - \, \sigma \, \gamma
\biggr)\frac{1 + k^2}{\beta} \biggr]\,\frac{\partial
\phi_1^{(1)}}{\partial \zeta}\,
\, , \nonumber \\
&\equiv &
 \, i \, c^{(21)}_5
\,\frac{\partial \phi_1^{(1)}}{\partial \zeta} \, .\qquad
 \label{coeffs21}
\end{eqnarray}
The choice of the value of $\tilde A$ is arbitrary; we shall take
$\tilde A = 0$.

The equations for $n = 2$, $l = 2$ provide the amplitudes of the
second order harmonics, which are found to be proportional to the square of
the corresponding $S_1^{(1)}$ elements e.g.
in terms of $\phi_1^{(1)}$
\begin{eqnarray}
n_2^{(2)} \,  &= & \, \biggl[\frac{1}{\omega} A \, + \, \frac{(1 +
k^2)^2}{\beta^2}\biggr] \, \equiv  \, c^{(22)}_1
\,{\phi_1^{(1)}}^2 \, ,\nonumber
\\
\mathbf{k\cdot u}_2^{(2)} \,  &= & \, \frac{(1 + k^2) \, \omega
}{6 \beta^3 k^2} \, \biggl[ 2\,s\, \alpha\, \beta^2 \, + \, 3
\beta \,(1 + k^2) (1 + 2 k^2) + \, 2\,\gamma^2 \, \sigma \, (1 +
k^2)^2 \, (1 + 4 k^2) \biggr] \, {\phi_1^{(1)}}^2\, \nonumber
\\
& \equiv  & \, A \,{\phi_1^{(1)}}^2 \nonumber \,= c^{(22)}_2
\,{\phi_1^{(1)}}^2 \, ,\nonumber
\\
p_2^{(2)} \,  &= & \,  \gamma \, \biggl[\frac{1}{\omega} A \, +
\gamma\, \frac{(1 + k^2)^2}{\beta^2}\biggr] \, \equiv  \,
c^{(22)}_3 \,{\phi_1^{(1)}}^2 \, ,\nonumber
\end{eqnarray}
and
\begin{equation}
\phi_2^{(2)} \,  = \, \frac{1}{4 k^2 + 1} \, \biggl\{ s \,
\beta \, \biggl[ \frac{1}{\omega} A \, + \, \frac{(1 +
k^2)^2}{\beta^2}\biggr] + \, \alpha \biggr\} \, {\phi_1^{(1)}}^2\,
\equiv
\, c^{(22)}_4 \,{\phi_1^{(1)}}^2 \, .
 \label{coeffs22}
\end{equation}
Notice that these expressions are {\em isotropic} i.e. independent of
the value of $\theta$.

The nonlinear self-interaction of the carrier wave also results in the
creation of a zeroth harmonic, in this order; its strength is analytically
determined by taking into account the $l = 0$ component of the three first
third-order reduced equations (i.e. (\ref{geneqn1}) - (\ref{geneqn3}) for
$n = 3$, $l = 0$) together with the corresponding fourth 2nd-order equation
(i.e. (\ref{geneqn4}) for $n = 2$, $l = 0$).  The result is conveniently
expressed in terms of the square modulus of the ($n = 1$, $l = 1$) quantities,
e.g. in terms of $|\phi_1^{(1)}|^2  = (\phi_1^{(1)})^*\, \phi_1^{(1)}$
\begin{eqnarray}
n_0^{(2)} \,  &= & \, \frac{-1}{\beta + \gamma \sigma - v_g^2}\,
\frac{1}{\beta}\, \biggl[  1 + 2 s \alpha \beta \, + k^2 \,+ 2 \,
\cos^2\theta
\nonumber \\
&&
\qquad \qquad \qquad \qquad \qquad + \,  \gamma \, \sigma \,
\frac{(1 + k^2)^2}{\beta}\, (\gamma  + \,2 \cos^2\theta -1)\biggr]
|\phi_1^{(1)}|^2 \, \nonumber \\
& \equiv &\, B\, |\phi_1^{(1)}|^2
 \, = \, c^{(20)}_1 \,|\phi_1^{(1)}|^2 \, , \nonumber
\\
\mathbf{k\cdot u}_0^{(2)} \,  & = & \, \frac{-1}{\beta + \gamma
\sigma - v_g^2}\, \frac{\cos\theta}{\beta^2}\, \biggl\{  2
\omega\, (\beta +  \gamma \sigma) (1+k^2)^2\,\cos\theta \nonumber
\\ &&
\qquad \qquad \qquad \qquad \qquad \qquad + k\, v_g\, \bigl[\beta
\,(1 + k^2 \,+ 2 s \alpha \beta) \, + \sigma \,\gamma (\gamma - 1)
(1+k^2)^2 \bigr] \biggr\} \nonumber
\\
& \equiv  & \, c^{(20)}_2 \,|\phi_1^{(1)}|^2 \, ,\nonumber
\\
p_0^{(2)} \,  &= & \,\gamma \,\biggl[ B \, + \, \frac{1}{\beta^2}
\, (\gamma - 1)\, (1 + k^2)^2) \biggr]  \,|\phi_1^{(1)}|^2 \,
\equiv  \, c^{(20)}_3 \,|\phi_1^{(1)}|^2 \, ,\nonumber
\\
\phi_0^{(2)} \,  &= & \,(s \,\beta \,B \, + 2 \, \alpha)
\,|\phi_1^{(1)}|^2 \, \equiv \, c^{(20)}_4 \,|\phi_1^{(1)}|^2 \, ,
 \label{coeffs20}
\end{eqnarray}
and
\begin{equation}
u_{0, x}^{(2)} \, = \,\biggl[ v_g\, B \, - 2 \,
 \frac{\omega \,(1 + k^2)^2}{\beta^2 \, k}\,
 \cos\theta \biggr] \,|\phi_1^{(1)}|^2 \,
\equiv  \, c^{(20)}_5 \,|\phi_1^{(1)}|^2 \, . \label{coeff205}
\end{equation}
It is expected, and indeed verified by a tedious yet
straightforward calculation, that upon setting $\sigma = 0$, $s =
-1$ in expressions (\ref{coeffs22}) and (\ref{coeffs20}), one
recovers exactly Eqs. (44) - (49) in Ref. \cite{AMS} [given (42) therein].

Notice, for rigor, that for `vanishing obliqueness' i.e. if
${\theta \rightarrow 0}$, one obviously has \(\mathbf{k\cdot u}_l^{(n)}
\,\rightarrow k \, u_l^{(n)}\) (by definition),
implying the condition:
$c_2^{(nl)} \,\rightarrow k\, c_5^{(nl)}$ (for ${\theta \rightarrow 0}$)
which is indeed
satisfied for all $n$, $l$, by the above formulae.

\subsection{Derivation of the Nonlinear Schr\"odinger Equation}

Proceeding to the third order in $\epsilon$ ($n=3$), the equation for
$l = 1$ yields an explicit compatibility condition to be imposed on
the right-hand side of the evolution equations which,
given the expressions derived previously,
can be cast into the form
\begin{equation}
A_1\, \frac{d \psi}{d\tau} + i\, A_2\, \frac{d^2 \psi}{d\zeta^2} +
\, i\, A_3\, |\psi|^2\,\psi = 0
\, ,
\label{NLSE1}
\end{equation}
where $\psi\, \equiv \,  \phi_1^{(1)}$ denotes the
amplitude of the first-order electric
potential perturbation; coefficients $A_{1, 2 ,3}$ are to be defined.
Now, multiplying by $i\, A_1^{-1}$, we obtain the familiar form of the
Nonlinear Schr\"odinger Equation
\begin{equation}
i\, \frac{\partial \psi}{\partial \tau} + P\, \frac{\partial^2
\psi}{\partial \zeta^2} + Q \, |\psi|^2\,\psi = 0 \, .
\label{NLSE}
\end{equation}
Recall that the `slow' variables $\{ \zeta, \tau \}$ were defined in
(\ref{slowvar}).

The {\em dispersion coefficient} $P = -A_2/A_1$ is related to the
curvature of the dispersion curve as \( P \,  = \, \frac{1}{2} \,
\frac{\partial^2 \omega}{\partial k_x^2} \,= \, \frac{1}{2}\,
\biggl[ \omega''(k) \, \cos^2\theta \, + \omega'(k) \,
\frac{\sin^2\theta}{k} \biggr] \); the exact form of P reads
\begin{equation}
P(k) \,  =\, \frac{1}{\beta}
\frac{1}{2\,\omega} \, \biggl(\frac{\omega}{k}\biggr)^4\,
\biggl[ \nu_1 - (\nu_1 + 3\, \frac{\nu_2}{\beta} \,\omega^2)\, \cos^2\theta
\biggr] \, ,
\label{Pcoeff}
\end{equation}
where we have defined
\begin{equation}
\nu_2 = \beta^3 \,\frac{3 \beta +
\gamma \sigma (3 - k^2)(1 + k^2)
}
{3 \,[\beta +
\gamma \,\sigma \,(1 + k^2)]^4 } \, .
\label{defnu2}
\end{equation}
Note that, just like $\nu_1$ defined above, $\nu_2 \rightarrow 1$
when $\sigma \rightarrow 0$; see that relation (51) in Ref.
                \cite{AMS}
is recovered from (\ref{Pcoeff}) in this case. If, furthermore, we
set $\beta = 1$ (in addition to $\sigma = 0$) in all expressions
describing our dispersion law i.e. (\ref{dispersion}), (\ref{vg}),
(\ref{Pcoeff}) above, we obtain respectively (3), (11), (4) in Ref.
\cite{Kako}.

It seems appropriate, here, to point out the qualitative difference
between $P$ given in (\ref{Pcoeff}) as compared to relevant
previous expressions: the existence of $\sigma$ may
affect the sign of the $P$ coefficient.
For instance, taking
$\sigma = 0$ (i.e. $\nu_1 = \nu_2 = 1$), $P$ is readily seen to be
negative for parallel modulation, i.e. setting $\theta = 0$;
however, for $\sigma \ne 0$ this is no longer the case, since $P$
changes sign at some critical value of k (to see this, study the
sign of $\nu_2$ versus $k$ \cite{commentsignP}).
Furthermore, a similar remark holds
for the effect of an oblique modulation on the sign of $P$; we will come back
to this subtle point in the next subsection.

The {\em nonlinearity coefficient} $Q = - A_3/A_1$ is due to
the carrier wave self-interaction.
Distinguishing different contributions,
$Q$ can be split into five distinct parts, viz.
\begin{equation}
Q = \, Q_0 \, +\, Q_1 \, +\, Q_2 \, +\, Q_3 \, +\, Q_4  \, ,
\label{Qstructure}
\end{equation}
reflecting the similar structure of $A_3$
\begin{equation}
A_3 = \, A_3^{(0)} \, +\, A_3^{(1)} \, +\, A_3^{(2)} \,
+\, A_3^{(3)} \, +\, A_3^{(4)} \, .
\label{A3structure}
\end{equation}
In order to trace the influence of the various parameters,
let us define all
quantities in full detail. First, $A_3^{(0)}$ (as well as $Q_0 = -
A_3^{(0)}/A_1$) is related to the self-interaction due to the zeroth
harmonic, viz.
\begin{equation}
A_3^{(0)} = \, - \beta \, k^2\, (c_1^{(11)} c_2^{(20)} \, +\,
c_2^{(11)} c_1^{(20)})
 \, - \, s \, \omega\, 2 \alpha \, k^2\, c_4^{(11)} c_4^{(20)}
  \, - \,\omega\, (1 + k^2)\, c_2^{(11)} c_2^{(20)}  \, ,
\label{A30coeff}
\end{equation}
while $A_3^{(2)}$ (related to $Q_2 = - A_3^{(2)}/A_1$) is
the analogue quantity due to the second harmonic
\begin{equation}
A_3^{(2)} = \, - \beta \, k^2\, (c_1^{(11)} c_2^{(22)} \, +\,
c_2^{(11)} c_1^{(22)})
 \, - \, s \, \omega\, 2 \alpha \, k^2\, c_4^{(11)} c_4^{(22)}
   \, - \,\omega\, (1 + k^2)\, c_2^{(11)} c_2^{(22)}
\, .
\label{A32coeff}
\end{equation}
All coefficients $c_j^{(nl)}$ were defined previously.
Now, $Q_1 = - A_3^{(1)}/A_1$ is
simply the nonlinearity contribution from the cubic term in
(\ref{reducedeqs}d) (often omitted in the past)
\begin{equation}
A_3^{(1)} =
 \, + 3 \, s \, \alpha' \, \omega  \, {(c_4^{(11)})}^{3}\, k^2 \, ,
\label{A31coeff}
\end{equation}
Finally, $A_3^{(3)}$ (related to $Q_3 = - A_3^{(3)}/A_1$) is
the ($\sigma$- related) result of the third line in (\ref{geneqn2})
\begin{equation}
A_3^{(3)} = \, - \sigma\, k^2\, (1 + k^2) \,
\biggl[ \gamma \, c_2^{(11)} \, (c_3^{(20)} - \, c_3^{(22)}) \,
+ \, 2 \gamma \,c_3^{(11)} c_2^{(22)}
+ \, c_3^{(11)} \, (c_2^{(20)} - \, c_2^{(22)}) \,
+ \, 2 \,c_2^{(11)} c_3^{(22)}
\biggr] \, ,
\label{A33coeff}
\end{equation}
while $A_3^{(4)}$ (and $Q_4 = - A_3^{(4)}/A_1$) is
due to the last two lines in (\ref{geneqn2})
\begin{equation}
A_3^{(4)} = \, - \omega\, (1 + k^2) \,
\biggl[
(\omega\, c_2^{(11)} - s k^2 \, c_4^{(11)})\,
(c_1^{(22)} - \, c_1^{(20)}) \,
- \, 2 \,c_1^{(11)} (\omega\, c_2^{(22)} - s k^2 \, c_4^{(22)})\,
\,
+ \,(c_2^{(11)})^2 c_1^{(11)}
\biggr] \, .
\label{A34coeff}
\end{equation}
We note that
$A_1$ is everywhere defined as
\begin{equation}
A_1 = \, - s \, \frac{2}{\beta}\, (1 + k^2)^2\, \omega^2 \, ,
\label{A1coeff}
\end{equation}
i.e. by using (\ref{dispersion})
\begin{equation}
A_1^{-1} = \, - s \, \frac{1}{2 \,\beta}\,\frac{1}{\omega^2}\,
\biggl( \frac{\omega^2}{k^2}\, - \gamma \, \sigma \biggr)^2
\label{A1coeff1}
\end{equation}
(reducing to:
$A_1^{-1} = \, - s \, \frac{1}{2 \,\beta}\,\frac{\omega^2}{k^4}$ for
$\sigma = 0$).
Remember that $Q_3$ and $Q_4$ are plainly absent from the previous
results in Ref. \cite{AMS} (i.e. for $\sigma = 0$) and so is, in fact,
$Q_1$.

Substituting from the expressions derived above for the
coefficients $c_j^{(nl)}$ and re-arranging, we obtain
\begin{eqnarray}
Q_0 &=& \, + \, \frac{1}{2 \omega}\,\frac{1}{\beta^2}\,
\frac{1}{(1 + k^2)^2}\,
\frac{1}{\beta + \gamma \sigma - v_g^2}\,\times
\nonumber \\ &&
\biggl\{  \beta \, k^2 \,
                        \biggl[
    \,\beta \,
        \bigl[ 3 + 6 k^2 + 4 k^4 + k^6
                        + 2 \,\alpha \,\beta \bigl(s \,(2 k^2 + 3) + \,2 \,\alpha \,v_g^2
                \bigr) \bigr]
\nonumber \\ &&
    \qquad \qquad + \, \gamma \, \sigma \,
            \bigl[ \,(\gamma + 1)\, (1 + k^2)^3
        + \,2 \,\alpha \,\beta \,\bigl(-2 \alpha \beta + s\, \gamma \, (1 + k^2)^2
        \bigr) \bigr]
\nonumber \\ &&
        \qquad \qquad + \,
                      \bigl[
                     \beta \, (2 + 4 k^2 + 3 k^4 + k^6 + 2 s \alpha \beta )
                 + 2 \gamma  \, \sigma \, (1 + k^2)^2 \, (1 + k^2 + s
                 \alpha \beta)
              \bigr] \, \cos 2\theta \biggr]
\nonumber \\ &&
      + \, 2 \, (1 + k^2)^4 \,(\beta + \gamma \sigma)\, \omega^2 \, \cos^2\theta
\nonumber \\ &&
       + \, k \, (1 + k^2) \, \biggl[\beta k^2 + \omega^2 \, (1 + k^2) \biggr]\,
       \frac{v_g}{\omega} \, \times
\nonumber \\ &&
       \qquad \qquad \qquad \qquad \qquad
       \biggl[ \beta \, (1+ k^2 + 2 s \alpha \beta)\,
       + \, \gamma \, (\gamma - 1)\, \sigma \, (1+k^2)^2 \biggr]    \,
       \cos\theta \biggr\}
\, ,
\label{Q0coeff}
\\ Q_1 &=&
\, \frac{3 \, \alpha' \beta}{2 \,\omega}\, \frac{k^2}{(1 + k^2)^2}
\, ,\label{Q1coeff}
\\
Q_2 &= & \, - \, \frac{1}{12 \,\beta^3}\,\frac{1}{\omega}\,
 \frac{1}{k^2 \, (1 + k^2)^2} \,\times
\nonumber \\ &&
 \biggl\{  2 \beta \, k^2 \,
                  \biggl[ 5 \,s \,\alpha \,\beta^2 \, (1 + k^2)^2 +
                  \, 2 \alpha^2 \beta^3
              + \,2 \,\gamma^2 \, \sigma \,  (1 + k^2)^4
              \, (1 + 4 k^2)
\nonumber \\ &&
\qquad \qquad  \qquad \qquad    + \,\beta \, (1 + k^2)^3 \, (3 + 9 k^2
+ 2 \,s \,\alpha \,\gamma^2
              \,\sigma) \biggr]
\nonumber \\ &&
\qquad + \, (1 + k^2)^3 \, \omega^2 \, \biggl[ \beta \, (3 + 9 k^2 + 6 k^4 + 2 s \alpha \beta)
\,
+\, 2 \, \gamma^2 \, \sigma \, (1 + k^2)^2 \, (1 + 4 k^2)
\biggr]
\biggr\}
\, .
 \label{Q2coeff}
\end{eqnarray}
Finally, the coefficients $Q_3 = -
A_3^{(3)}/A_1$ and $Q_4 = -
A_3^{(4)}/A_1$ can be directly computed
from (\ref{A33coeff}) - (\ref{A1coeff}) above; the lengthy final
expressions are omitted here.

Once substituted in (\ref{Qstructure}), these expressions
provide the final expression for the nonlinearity coefficient $Q$.
One may readily check, yet after a tedious calculation, that expressions
(\ref{Q0coeff}) and (\ref{Q2coeff}) reduce to (53) and (54) in Ref. \cite{AMS}
for $\sigma = 0$. However, the remaining coefficients $Q_1$, $Q_3$, $Q_4$
were absent in all previous studies of the DA waves, to the best of our knowledge.
Their importance will be discussed in the following.
Note that $Q_1$, $Q_2$ do not depend on the angle $\theta$.

\subsection{Behaviour of coefficients for small $k$}

A preliminary result regarding the behaviour (and the sign) of the
NLSE coefficients $P$ and $Q$, at least for long wavelengths, may
be obtained by considering the limit of small $k \ll 1$ in the
above formulae.

The parallel ($\theta = 0$) and oblique ($\theta \ne 0$) modulation cases
have to be distinguished straightaway.
For small values of $k$ ($k \ll 1$), $P$ is negative and
varies as
\begin{equation}
P \bigr|_{\theta=0} \,  \approx - \frac{3}{2} \,
\frac{\beta}{\sqrt{\beta + \gamma
\sigma}} \,k
\end{equation}
in the parallel modulation case (i.e. $\theta = 0$), thus
tending to zero for vanishing $k$, while for $\theta \ne 0$, $P$
is positive and goes to infinity as
\begin{equation}
P \bigr|_{\theta\ne0} \,  \approx \frac{\sqrt{\beta +
\gamma \sigma}}{2 \, k} \, \sin^2\theta
\end{equation}
for vanishing $k$.
Therefore, the slightest deviation by $\theta$ of the amplitude variation direction
with respect to the wave propagation direction results in a change in sign of the
dispersion coefficient $P$.
Given the importance of the coefficient product $P Q$ (to be
discussed in the
next Section),
one may wonder whether this is sufficient for
the stability
characteristics of the DA wave to change.
Let us see what happens with the $Q$ in the limit of small $k$.

For all cases, $Q$ varies as $\sim 1/k$ for small $k \ll 1$ \cite{commentlowQ};
the exact expression in fact depends on the angle $\theta$.
In the general case ($\theta \ne 0$),
the result reads
\begin{equation}
Q \bigr|_{\theta\ne 0} \, \approx \, - \frac{1}{12 \,\beta^3} \,
\frac{1}{\sqrt{\beta + \gamma \sigma}}\,
 [\beta \, (2 s \alpha \beta  + 3) + 2 \gamma^2 \sigma ] \,
[\beta \, (2 s \alpha \beta  + 3) + \gamma \,  (\gamma  + 1)\,\sigma ]\,
 \frac{1}{k} \, .
 \label{lowQthetageneral}
 \end{equation}
A careful study shows that $Q$ is negative, in fact,
for all possible values of the physical  parameters of interest
(i.e. $\alpha$,  $\beta$, $\gamma$,  $\sigma$ - all positive -
{\em and} $s \pm 1$).
For vanishing $\theta$, however,
the approximate expression for $Q$, yet apparently quite similar,
is now {\em positive}, i.e.
\begin{equation}
Q \bigr|_{\theta = 0} \, \approx \, + \frac{1}{12 \,\beta^3} \,
\frac{1}{\sqrt{\beta + \gamma \sigma}}\,
 [\beta \, (2 s \alpha \beta  + 3) + 2 \gamma \sigma ] \,
[\beta \, (2 s \alpha \beta  + 3) + \gamma \,  (\gamma  + 1)\,\sigma ]\,
 \frac{1}{k} \, .
  \label{lowQthetazero}
 \end{equation}

In conclusion, both coefficients $P$ and $Q$ change sign when
`switching on' \textsl{theta}. Indeed, obliqueness in modulation is
expected to influence the stability profile of the system; this
point seems to confirm (and complete) the general qualitative
arguments put forward in Ref. \cite{Kako} for the ion acoustic wave
in an electron ion plasma without dust. Nevertheless, at all cases,
the product of $P$ and $Q$ is negative for small $k$,
ensuring, as we shall see in the following section, stability for
long perturbation wavelengths. As a by-product of this analysis,
we see that taking into account $Q_1$, $Q_3$ and $Q_4$
does not seem to influence the dynamics
in the low wavenumber $k$ parameter range.

\section{Stability analysis}

The standard stability analysis [8, 34]
consists in linearizing around the
monochromatic (Stokes's wave) solution of the NLSE (\ref{NLSE})
\[\psi \, = \, {\hat \psi} \, e^{i Q |\psi|^2 \tau} \, + \, c.c. \, ,
\]
(notice the amplitude dependence of the frequency) by setting
\[{\hat \psi} \, = \, {\hat \psi}_0 \, + \, \epsilon \, {\hat \psi}_1 \, ,
\]
and taking the perturbation ${\hat \psi}_1$ to be of the form:
${\hat \psi}_1 \, = \,
{\hat \psi}_{1, 0} \,e^{i ({\hat k} \zeta - {\hat \omega} \tau)} \, + \, c.c.$,
(the perturbation wavenumber $\hat k$ and the frequency $\hat \omega$
should be distinguished from their carrier wave homologue quantities,
denoted by $k$ and $\omega$).
Now, substituting into (\ref{NLSE}), one readily obtains the nonlinear
dispersion relation
\begin{equation}
\hat \omega^2 \, = \, P^2 \, \hat k^2 \, \biggl(\hat k^2 \, - \, 2 \frac{Q}{P}
|\psi_{1, 0}|^2 \biggr) \, .
\end{equation}
One immediately sees that the wave will be {\em stable} for all
values of $\hat k$ if the product $P  Q$ is negative. However, for
positive $P  Q > 0$, instability sets in for wavenumbers below a
critical value $\hat k_{cr} = \sqrt{2 \frac{Q}{P}} |\hat \psi_{1,
0}|$, i.e. for wavelengths above a threshold: $\lambda_{cr} = 2
\pi/\hat k_{cr}$; defining the instability growth rate \( \sigma =
|Im\hat\omega(\hat k)| \), we see that it reaches its maximum
value for $\hat k = \hat k_{cr}/\sqrt{2}$, viz.
\begin{equation} \sigma_{max} =
|Im\hat\omega|_{\hat k = \hat k_{cr}/\sqrt{2}} \,=\, | Q |\, |\hat
\psi_{1, 0}|^2  \, . \label{growthrate}
\end{equation}
In brief, we see that the instability condition depends only
on the sign of the product $P Q$, which can now be studied numerically,
relying on the exact expressions derived in the preceding Section.

In the contour plots presented below (see figures \ref{figure1},
\ref{figure2}, \ref{figure5}), we have depicted the $P Q = 0$
boundary curve against the normalized wavenumber $k/k_D$ (in
abscissa) and angle $\theta$ (between $0$ and $\pi$); the area in
black (white) represents the region in the $(k - \theta)$ plane
where the product is negative (positive); instability therefore
occurs for values inside the {\em white} area. We have considered
values of the wavenumber $k$ between zero and upto 4 times the
Debye wavenumber $k_D$ (yet mostly focusing our attention on the
low $k$ region). Pitch angle $\theta$ is allowed to vary between
zero and ${\pi}/{2}$; as a matter of fact, all plots are
$\frac{\pi}{2}$- periodic, i.e. symmetric upon reflection with
respect to either the $\theta = 0$ or the $\theta = \frac{\pi}{2}$
lines. We have chosen a fixed set of representative values:
$\alpha = 5\cdot 10^{-3}$, $\alpha' = 2 \alpha^2/3 \approx
1.6\cdot 10^{-5}$ and $\beta \approx 100$, corresponding to
$Z_d/Z_i = 10^3$ and $T_e/T_i = 10$ (we have taken $\gamma = 2$,
$\sigma = 1$ for the plots).

For negative dust ($s = -1$; see fig. \ref{figure1}) the product
possesses positive values for angle values between zero and
$\theta \approx 51{}^\circ$; we see that instability sets in above
a wavenumber threshold which is clearly seen to decrease as the
modulation pitch angle $\theta$ increases from zero to
approximately 17 degrees, and then increases again up to $\theta
\approx 51{}^\circ$. Nevertheless, beyond that value (and up to
$\pi/2$) the wave remains stable; this is even true for the
wavenumber regions where the wave would be {\em unstable} to a
parallel modulation: see e.g. the interval where $\theta = 0$ and
$k/k_D \in [1.0, 3.6]$ approximately, in figure \ref{figure1}. The
inverse effect is also present: even though certain $k$ values
correspond to stability for $\theta = 0$, the same modes may
become unstable when subject to an oblique modulation ($\theta \ne
0$); this is mostly true for long wavelengths (small $k$). Notice
the periodicity with respect to $\theta$.

A similar behaviour is witnessed in the case of positive dust
($s = +1$; see fig. 2), yet the instability threshold $k_{cr}$ for
a given value of $\theta$ is quite higher: positive dust rather
appears to {\em favour} stability.

In all cases, the wave appears to be globally stable for large
angle $\theta$ modulation (between 0.9 and $\pi/2$ radians, i.e.
$51{}^\circ$ to $90{}^\circ)$ and unstable for smaller values of
$\theta$. For parallel modulation ($\theta = 0$), the sign of the
product $P Q$ is basically opposite to that of $Q$, since $P < 0$
for all values of $k$; the wave is then stable for large
wavelengths $\lambda \gg \lambda_D$ (i.e. for $k/k_D \ll 1$), and
potentially unstable for higher values of $k$ (a similar
qualitative behaviour has been reported for the ion-acoustic wave
case (i.e. without dust) [11 - 15].

A final word is in row, concerning the effect of taking into
account the pressure evolution equation (\ref{eqp}), often omitted
for simplicity. Given the above results, this amounts to wondering
what the difference would be, should we simply set $\sigma = 0$ in
expression (\ref{Pcoeff}) for $P$ and thus plainly omit $Q_3$ and
$Q_4$, defined above. A qualitative answer is attempted in figure
\ref{figure5}, where we have depicted the $P Q$ product in this
case. The qualitative results obtained so far do not seem to be
strongly modified, at least not for low values of $k$ (say, below
$k \approx 2 k_D$) and definitely not as far as the angle
dependence of stability is concerned. The difference in stability
regions obtained for higher $k$ is rather negligible for long
wavelengths (say, below $k \approx 1.5 k_D$). Nevertheless,
including the pressure equation in the description seems to
describe the problem in a more precise manner, and also somewhat
restricts the instability region, since stability is now predicted
for short wavelengths (above, say, $k \approx 3.6 k_D$) and low
$\theta$; compare figs. \ref{figure1}a, \ref{figure2}a to
\ref{figure5}a, \ref{figure6}a, respectively.

\section{Nonlinear excitations}

Let us discuss the possibility of the existence of localized
excitations in our system.
The NLSE (\ref{NLSE}) is known to possess distinct types of localized
constant profile (solitary wave) solutions, depending on the sign of the
product $P Q$.
We shall now briefly outline the method employed to derive their form
and discuss their relevance to our problem.

Following Ref. \cite{Fedele}, we may seek a solution of Eq. (\ref{NLSE})
in the form
\begin{equation}
\psi(\zeta, \tau) = \sqrt{\rho(\zeta, \tau)} \, e^{i\,\Theta(\zeta, \tau) }
\, ,
\label{ansatz}
\end{equation}
where $\rho$, $\sigma$ are real variables which are determined by
substituting into the NLSE and
separating real and imaginary parts.
The different types of solution thus obtained are clearly
summarized in the following
paragraphs.

\subsection{Bright solitons}

For $P Q > 0$ we find the {\em (bright) envelope soliton} \cite{commentFedele1}
\begin{equation}
\rho = \rho_0 \, sech^2\biggl(\frac{\zeta - u\, \tau}{L} \biggr)
\, , \qquad  \Theta =
\frac{1}{2 P} \, \bigl[ u\,\zeta \, - (\Omega + \frac{1}{2} u^2)\tau \bigr]
\, ,
\end{equation}
representing a localized pulse travelling at a speed $u$ and oscillating
at a frequency $\Omega$ (for $u = 0$). The pulse width $L$
depends on the (constant) maximum amplitude square $\rho_0$
as
\begin{equation}
L = \sqrt{\frac{2 P}{Q \,\rho_0}} \, .
\label{widthbright}
\end{equation}

\subsection{Dark solitons}

For $P Q < 0$ we have the {\em dark} envelope soliton ({\em hole}) \cite{commentFedele1}
\begin{eqnarray}
\rho & = & \rho_1 \, \biggl[ 1 - \, sech^2 \biggl(\frac{\zeta - u\, \tau}{L'}
\biggr)\biggr] \, = \, \rho_1 \,
 tanh^2 \biggl(\frac{\zeta - u\, \tau}{L'}
\biggr)
\, ,
\nonumber \\
\Theta & = & \frac{1}{2 P} \,
\biggl[ u\,\zeta \, - \biggl(\frac{1}{2} u^2 - 2 P Q \rho_1 \biggr) \,\tau \biggr]
\, ,
\label{darksoliton}
\end{eqnarray}
representing a localized region of negative wave density (shock)
travelling at a speed $u$. Again, the pulse width
depends on
the maximum amplitude square
$\rho_1$ via
\begin{equation}
L' = \sqrt{2 \biggl|\frac{P}{Q\,\rho_1}\biggr|} \, \qquad .
\label{widthdark}
\end{equation}

\subsection{Grey solitons}

It has been shown in Ref. \cite{Fedele} that looking for
velocity-dependent amplitude solutions, for $P Q < 0$,
one obtains the {\em grey} envelope solitary wave
\begin{eqnarray}
\rho & = & \rho_2 \, \biggl[ 1 - a^2\, sech^2 \biggl(\frac{\zeta - u\, \tau}{L''}
\biggr)\biggr] \, ,
\nonumber \\
\Theta & = & \frac{1}{2 P} \,
\biggl[ V_0\,\zeta \, - \biggl(\frac{1}{2} V_0^2 - 2 P Q \rho_2 \biggr)
\,\tau + \Theta_{10} \biggr]\,
- S
\, \sin^{-1} \frac{a\, \tanh\bigl(\frac{\zeta - u\, \tau}{L''}
\bigr)}{\biggr[  1 - a^2\, sech^2 \biggl(\frac{\zeta - u\, \tau}{L''}
\biggr) \biggr]^{1/2}} \, ,
\label{greysoliton}
\end{eqnarray}
which also represents a localized region of negative wave density;
$\Theta_{10}$ is a constant phase; $S$ denotes the product
$S = sign \,P \, \times sign \,(u - V_0)$.
In comparison to the dark soliton (\ref{darksoliton}), note that
apart from the maximum amplitude
$\rho_2$, which is now finite (i.e. non-zero) everywhere,
 the pulse width of this grey-type excitation
\begin{equation}
L'' = \sqrt{2 \biggl|\frac{P}{Q\,\rho_2}\biggr|}  \,\frac{1}{a}
\label{widthgrey}
\end{equation}
now
also depends on $a$, given by
\begin{equation}
a^2 \, = \, 1 \, + \, \frac{1}{2 P Q} \frac{1}{\rho_2} (u^2 - V_0^2) \, \le \, 1
\label{grey-depth}
\end{equation}
($P Q < 0$), an independent parameter representing the modulation depth
($0 < a \le 1$).
$V_0$ is an independent real constant which satisfies the condition \cite{Fedele}
\[
V_0 - \sqrt{2 |P Q|\, \rho_2} \, \le \, u \, \le \,V_0 + \sqrt{2 |P Q|\, \rho_2}
\quad ;
\]
for $V_0 = u$, we have $a = 1$ and thus recover
the {\em dark} soliton presented in the previous paragraph.

Summarizing, we see that the regions depicted in figs.
\ref{figure1}, \ref{figure2}, \ref{figure5}, \ref{figure6}
actually distinguish the regions where different types of
localized solutions may exist: bright (dark or grey) solitons will
occur in white (black) regions (the different types of NLS
excitations are exhaustively reviewed in \cite{Fedele}).
Furthermore, soliton characteristics will depend on the dispersion
laws via the $P$ and $Q$ coefficients; for instance, regions with
higher values of $P$ (or lower values of $Q$) - see figs.
\ref{figure3}, \ref{figure4} - will support wider (spatially more
extended) localized excitations.

\section{Conclusions}

This work has been devoted to the study
of the conditions for occurrence of the
modulational instability of the dust-acoustic waves propagating
in an unmagnetized dusty plasma.
Considering the Poisson-moment equations for the dust and allowing for
modulation to occur in an oblique manner, we have shown that the
DA wave modulational instability depends
strongly on the angle between the propagation and modulation directions.
As a matter of fact, the region of parameter values where
instability occurs is rather extended for angle $\theta$ values up to a certain threshold,
and, on the contrary, smeared out for higher $\theta$ values
(and up to 90 degrees, then going on in a $\frac{\pi}{2}$ - periodic fashion).

Furthermore, we have studied the possibility of the formation of localized
structures (solitary waves) in the system. Distinct types of
localized excitations (envelope solitons) have been shown to exist.
Their type and propagation characteristics depend on the carrier wave
wavenumber $k$
and the modulation angle $\theta$.

\begin{figure}[htb]
 \centering
 \resizebox{3in}{!}{
 \includegraphics[]{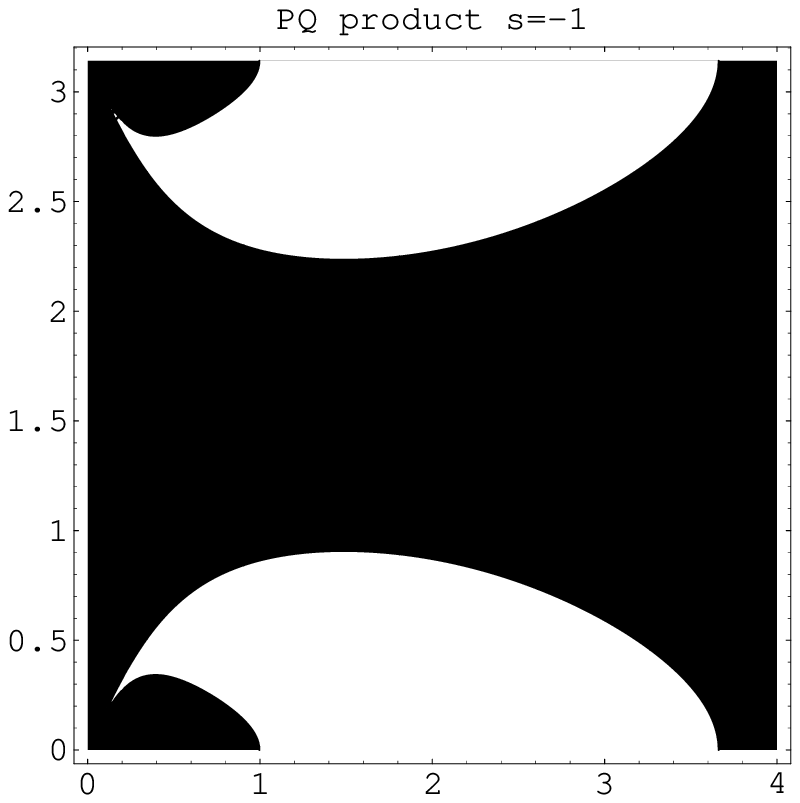}}
\vskip 1.5 cm \resizebox{3in}{!} {\includegraphics{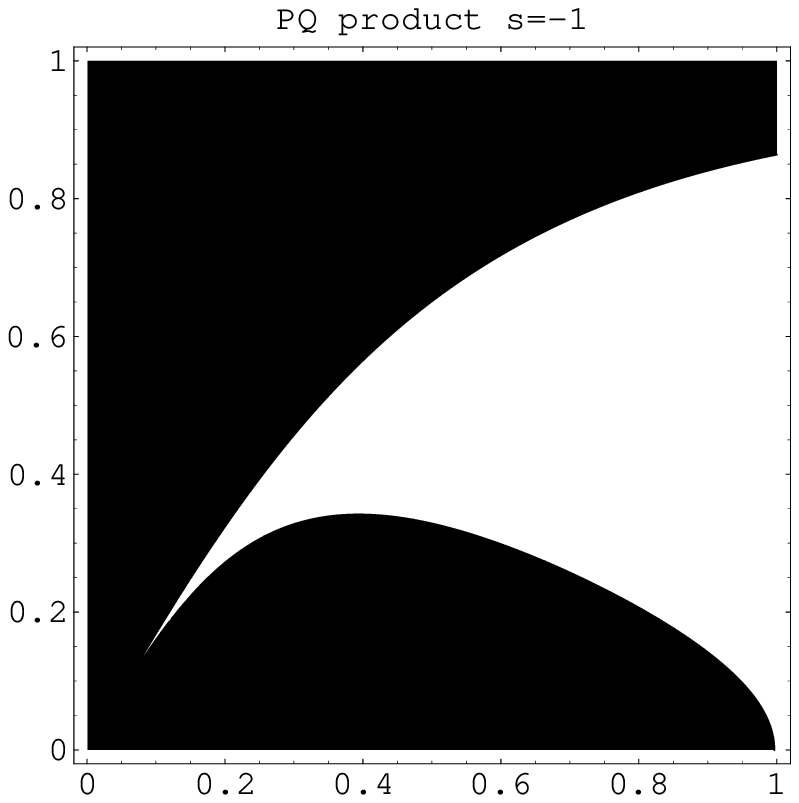} }
\caption{} \label{figure1}
\end{figure}

\begin{figure}[htb]
 \centering
 \resizebox{3in}{!}{
 \includegraphics[]{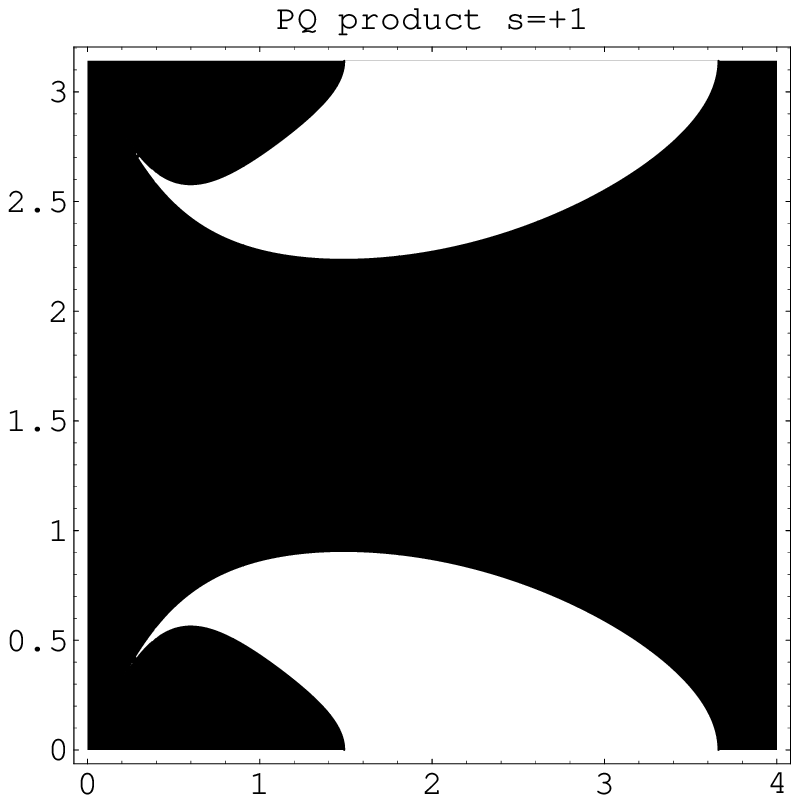}
} \vskip 1.5 cm
 \resizebox{3in}{!}{
\includegraphics{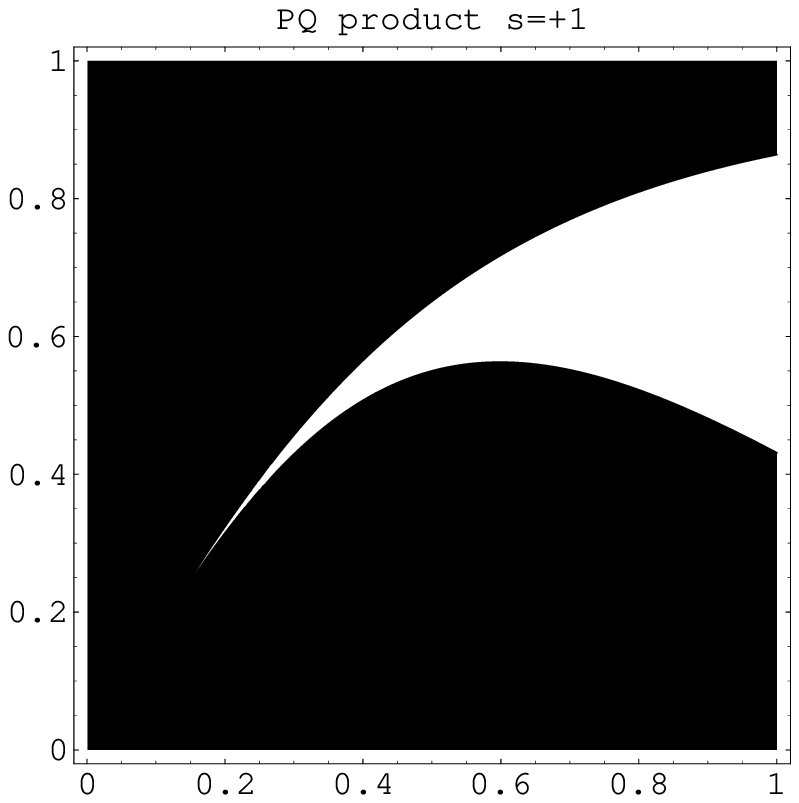}
} \caption{} \label{figure2}
\end{figure}

\begin{figure}[htb]
 \centering
 \resizebox{3in}{!}{
 \includegraphics[]{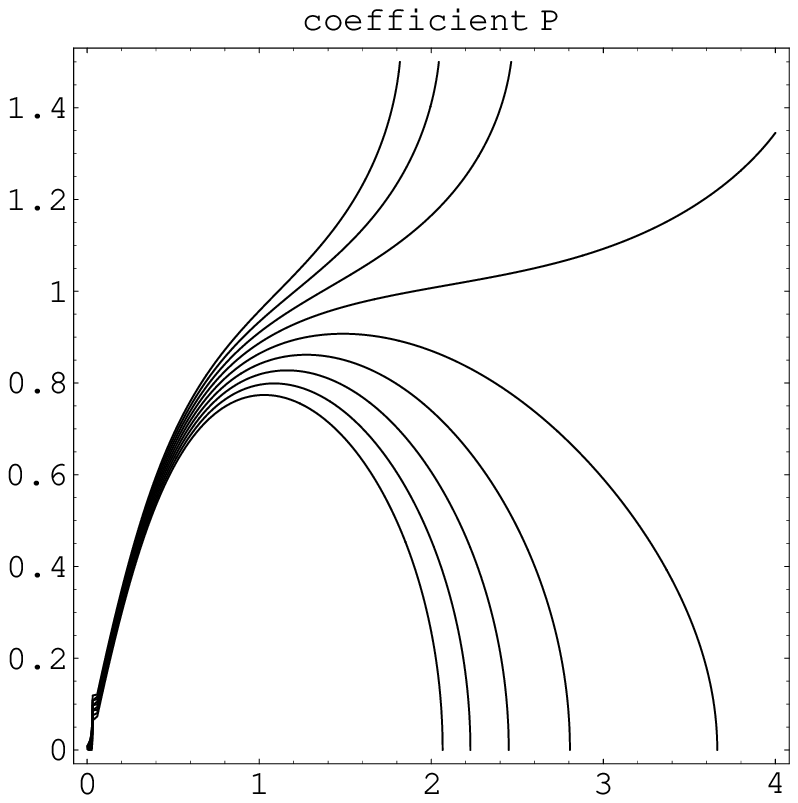}}
\vskip 1. cm
 \resizebox{3in}{!}{
\includegraphics[]{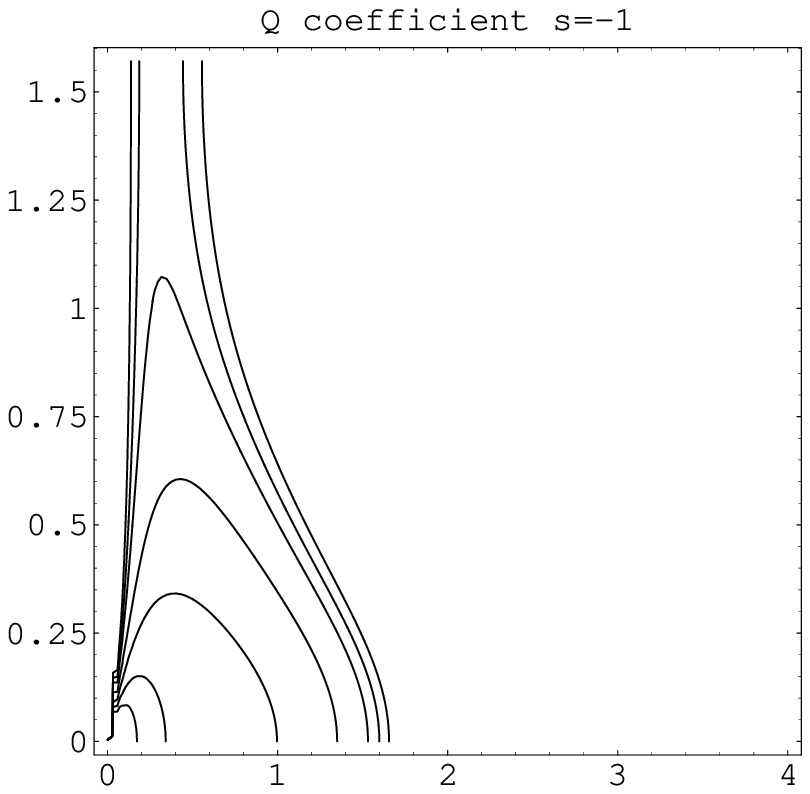}}
\vskip 1. cm
 \resizebox{3in}{!}{
\includegraphics[]{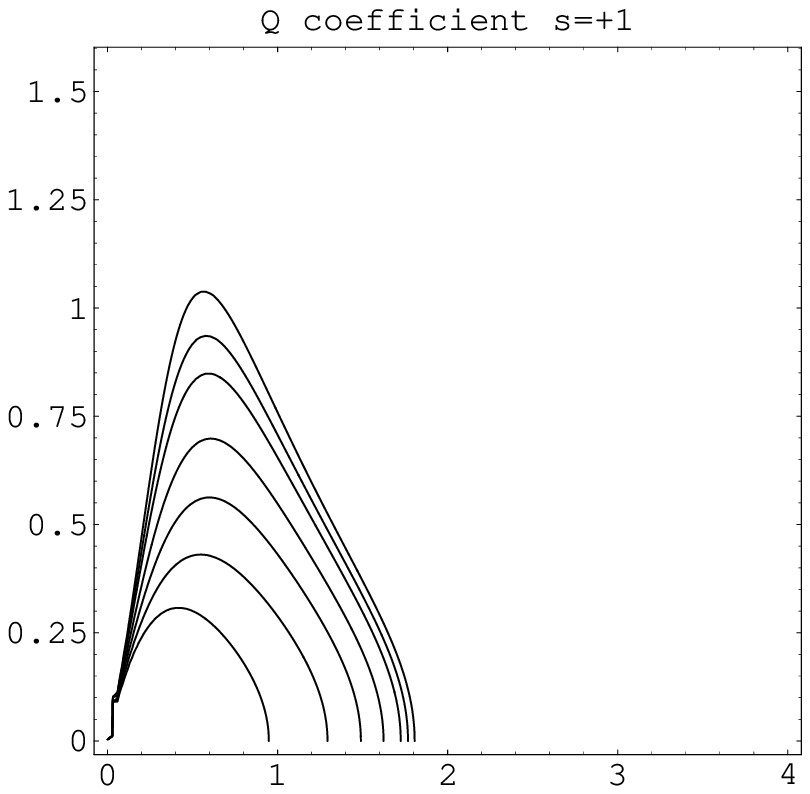}
} \caption{} \label{figure3}
\end{figure}

\begin{figure}[htb]
 \centering
 \resizebox{3in}{!}{
\includegraphics[]{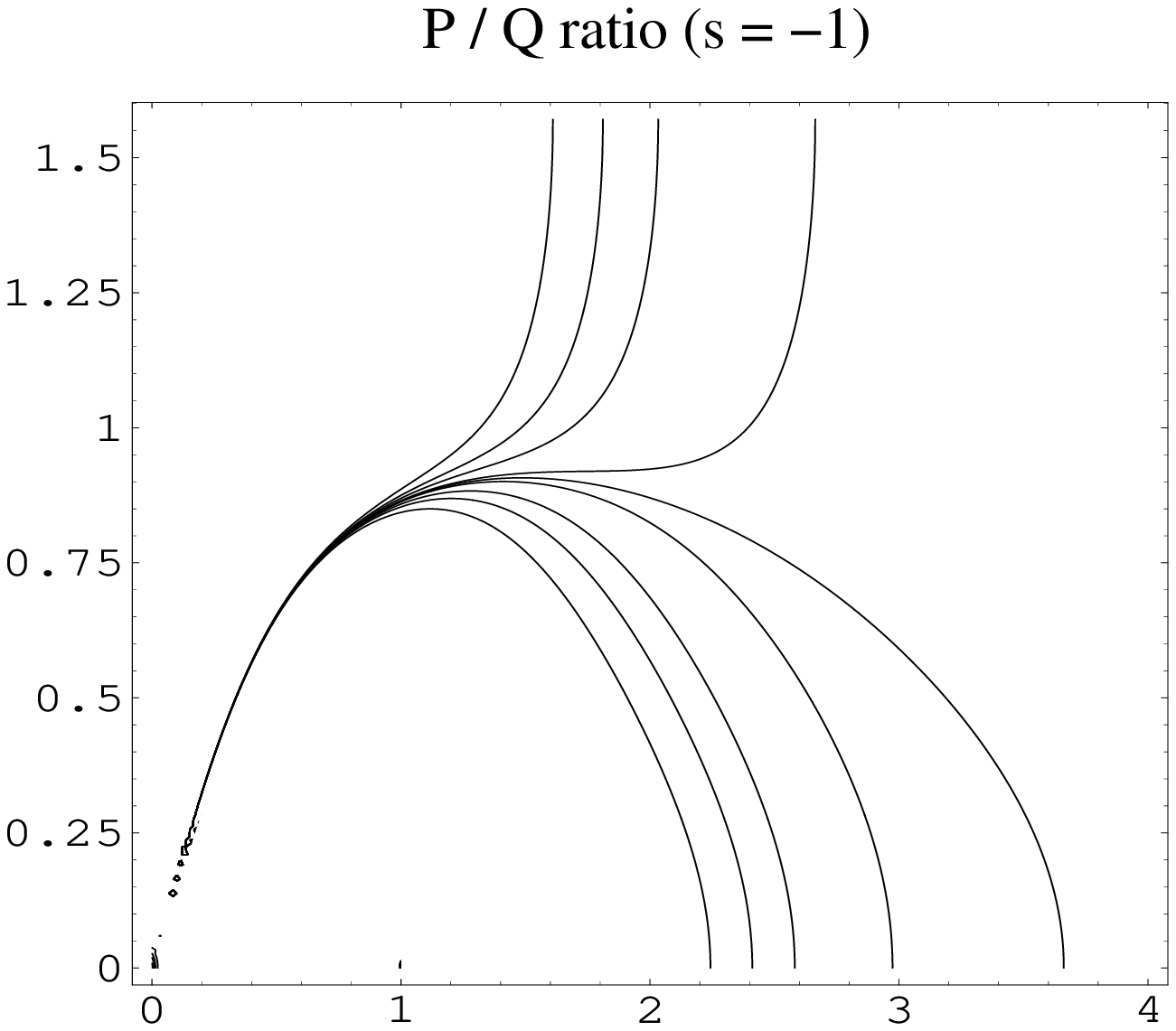}
} \caption{} \label{figure4}
\end{figure}

\begin{figure}[htb]
 \centering
 \resizebox{3in}{!}{
 \includegraphics[]{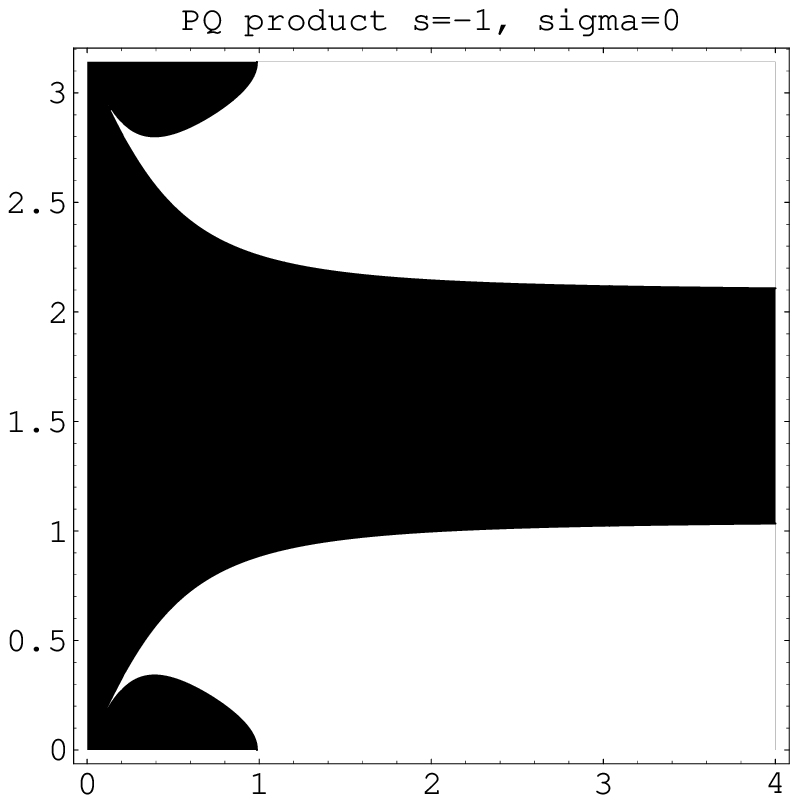}}
\vskip 1.5 cm
 \resizebox{3in}{!}{
 \includegraphics[]{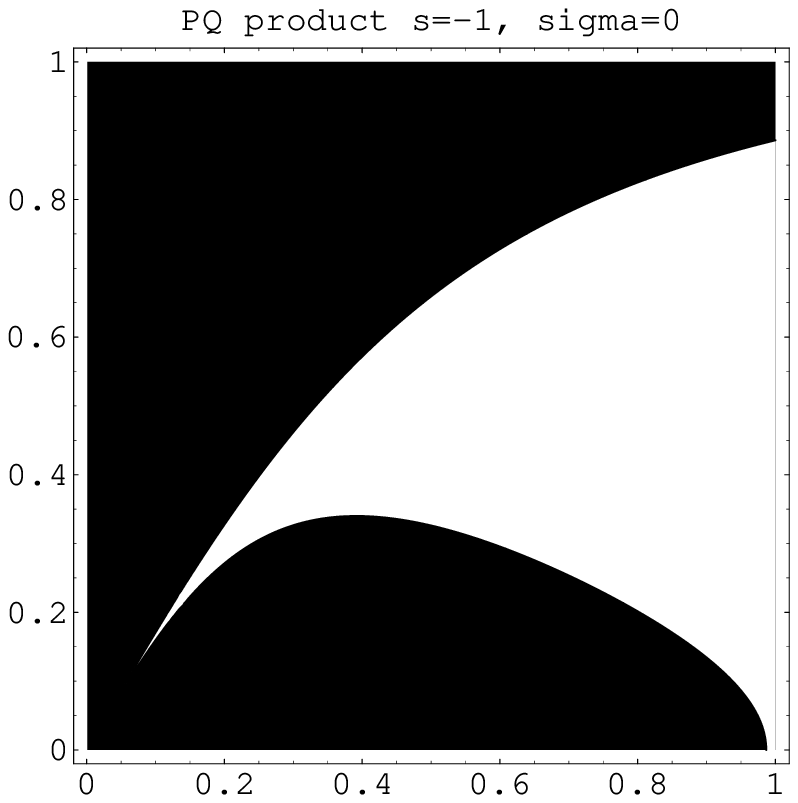}
} \caption{} \label{figure5}
\end{figure}

\begin{figure}[htb]
 \centering
 \resizebox{3in}{!}{
 \includegraphics[]{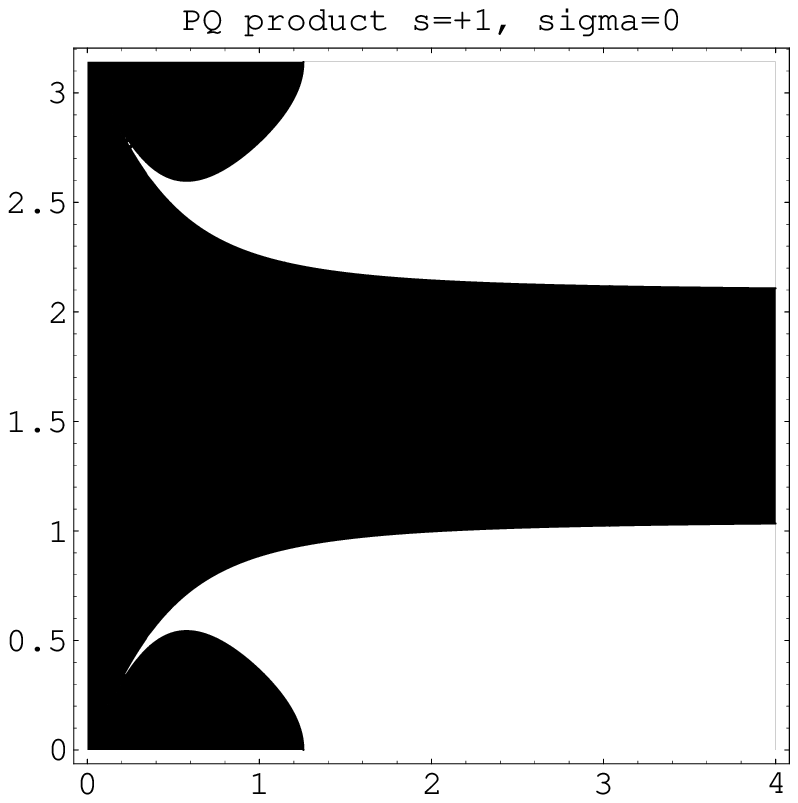}}
\vskip 1.5 cm
 \resizebox{3in}{!}{
\includegraphics[]{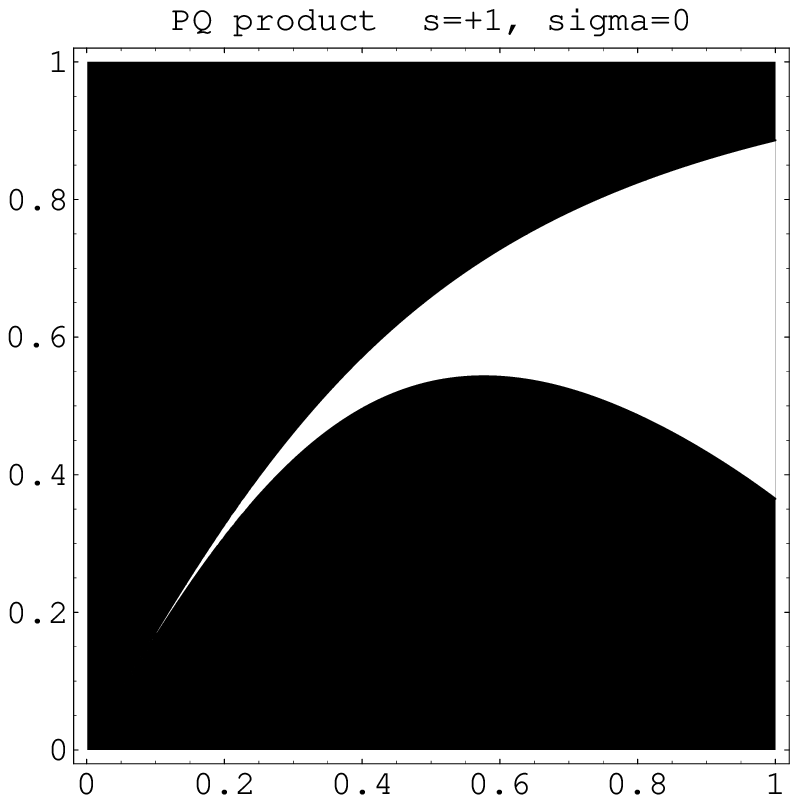}
} \caption{ } \label{figure6}
\end{figure}

Summarizing our results, we have seen that
\newline
(i) obliqueness in the amplitude modulation direction has a strong influence on
the conditions for the modulational instability to occur: regions which are
stable to a parallel modulation may become unstable when subject to an oblique
modulation, and vice versa;
\newline
(ii) large-angle modulation seems to have a stabilizing effect;
on the contrary, small-to-medium angle (say below 50 degrees) modulation
enhances instability;
\newline
(iii) DAW-related localized excitations may appear and propagate in a dusty
plasma;
modulationally stable (unstable) $(k, \theta)$ regions support envelope
solitary waves of the bright (dark) type;
\newline
(iv) the type and characteristics of the latter (localized modes)
depend on the value of $\theta$: for given low $k$, dark solitons
(or {\textsl{holes}}) are wider as $\theta$ becomes higher (see
fig. \ref{figure4}); for higher $k$, bright (dark) solitons become
narrower (wider) as $\theta$ increases; finally, for given
$\theta$ values below (above) a threshold of, say, 51 degrees,
bright (dark) excitations will be narrower (wider) for higher $k$
(see fig. \ref{figure4});
\newline
(v)
comparing the positive ($s= + 1$) to negative ($s = - 1$) dust cases,
we have shown that positive dust enhances stability
and rather favours
dark-type excitations (hole solitons);
furthermore, low $k$ dark envelope solitons appear to be narrower
with positive dust;
for higher $k$ there is practically no qualitative difference
between the two
dust charge sign cases;
\newline
(vi) As a final comment, let us point out that taking the dust
pressure equation (\ref{eqp}) into account, we have obtained a
wider stability region for small $\theta$ values, yet only for
high wavenumbers. The existence of dark-type localized envelopes
of high $k$ modes subject to slightly oblique (low $\theta$)
modulation is thus predicted; cf. figs. \ref{figure1}a,
\ref{figure2}a to \ref{figure5}a, \ref{figure6}a, respectively.
However, for wavenumbers below, say, $k = k_D$, there is no
qualitative difference due to the consideration of (\ref{eqp}).

Our aim has been to put forward a model study of the DAW modulation
which is generic, i.e. incorporating several previous descriptions,
which may be recovered for different choices of the physical
parameters involved in the formulation. Dust charge was assumed to be constant
and the plasma geometry was taken to be Cartesian and infinite, for simplicity.
Thus, our work
complements the investigation by Tang and Xue \cite{XueSept}
who examined only the modulational instability of DAWs against
oblique modulations, including an ad hoc charging equation and
a specific form of the adiabatic law for warm charged dust grains which
are negatively charged. The present paper, on the other hand, discusses the multi-
dimensional modulational instabilities of dust acoustic waves in plasmas containing both
negatively and positively charged dust grains, as well as provides a detailed
discussion of various types of dust acoustic envelope solitons
and their respective parameter regions of existence, leaving the choice of
the value of the parameter $\gamma = c_p/c_V$ free in the algebra.

\begin{acknowledgments}
This work was supported by the European Commission (Brussels)
through the Human Potential Research and Training Network
for carrying out the task of the project entitled: ``Complex
Plasmas: The Science of Laboratory Colloidal Plasmas and Mesospheric
Charged Aerosols'' through the Contract No. HPRN-CT-2000-00140.
\end{acknowledgments}

\newpage

\textbf{Figure captions}

\bigskip

Figure 1:

(a) The coefficient product $P Q = 0$ curve is represented against
normalized wavenumber $k/k_D$ (in abscissa) and angle $\theta$
(between $0$ and $\pi$); the area in black (white) represents the
region in the $(k - \theta)$ plane where the product is negative
(positive); instability therefore occurs for values inside the
white area. This plot refers to negative dust charge ($s = -1$).
(b) A close-up plot near the origin.

\bigskip

Figure 2:

Same as in figure 1, for positive dust charge ($s = +1$). Notice
that the stability region close to the origin gets narrower.
Positive dust charge seems to favour stability.

\bigskip

Figure 3:

(a) The curves for constant values (contours) of the dispersion
coefficient $P$ are represented against normalized wavenumber
$k/k_D$ (in abscissa) and angle $\theta$ (between $0$ and
$\pi/2$); In ascending order (from bottom to top), the curves
correspond to $P = -0.4, \, -0.3, ...,\, 0.3, \, 0.4$; $P$ clearly
increases with $\theta$, for a given wavenumber $k$. The
parameters used for this plot are as defined in fig. 1. (b) A
similar contour plot for the nonlinearity coefficient $Q$. In {\em
descending} order (from top to bottom), the curves correspond to
$Q = -0.003, \, -0.0025, \, -0.002, \, -0.002, \, 0, \, 0.001, \,
0.002$; $Q$ {\em decreases} with increasing $\theta$, in this
region. Remember that (the part of) these curves falling inside
the instability region (white sector in fig. \ref{figure1}) is
related to the instability growth rate $\sigma$ via
(\ref{growthrate}). Values of $\sigma/|\hat \psi_{1, 0}|^2$ above
a certain value are to be excluded, since they would fall inside
the stability (black) region: this element is absent in fig.
\ref{figure1}. This plot refers to negative dust charge ($s =
-1$). (c) The analogous contour plot (same values as in (b)) for
the nonlinearity coefficient $Q$ in the {\em positive} dust charge
($s = + 1$) case; $Q$ takes higher values here (cf. (b)), for
given $(k, \theta)$, leading to a more extended stability region
for large wavelengths ($k \ll k_D$).

\bigskip

Figure 4:

Contours of the ratio $P/Q$ (whose absolute value is related to
the soliton width; see (\ref{widthbright}), (\ref{widthdark})) are
represented against normalized wavenumber $k/k_D$ (in abscissa)
and angle $\theta$ (between $0$ and $\pi/2$); In descending order,
starting from above, the curves correspond to $P/Q = -20, \, -10,
\, -5, \, -1, \, 0, \, 1, \, 5, \, 10, \, 20$; the value of $P/Q$
decreases with $\theta$, for a given wavenumber $k$ above $k_D$,
so higher $\theta$ seems to favour narrower (wider) bright-
(dark-) type excitations. The same qualitative behaviour was
obtained for positive dust charge i.e. $s=+1$ (not depicted, for
the difference was unimportant).

\bigskip

Figure 5

The product $P Q$, as in fig. \ref{figure1}a, as results from the
pressure equation (\ref{eqp}) being omitted. Comparing to fig.
\ref{figure1}, notice that there is practically no qualitative
difference for low $k$ and for high $\theta$; however, predicted
behaviour changes above, say, $k \approx 1.5\, k_D$. This plot
refers to negative dust charge ($s = -1$).

\bigskip

Figure 6

Similar to \ref{figure5} but for $s = + 1$ (positive dust charge)
i.e. as in fig. \ref{figure2}, but omitting the pressure equation
(\ref{eqp}). Once more, the change in the qualitative analysis
does not appear to be dramatic.

\newpage

\textbf{Figures}

\newpage

\begin{thebibliography}{10}
\bibitem{comment2} Literally speaking, $\alpha$ may take negative values
for $\mu > \bigl(\frac{\alpha_2}{\alpha_1}\bigr)^2 =
\bigl( Z_i \,\frac{T_e}{T_i} \bigr)^2$, i.e.
if $s \frac{Z_d\, n_{d, 0}}{Z_i\, n_{i, 0}} > Z_i \,\frac{T_e}{T_i} - 1$,
indicating a very high concentration of {\em positive} dust in the plasma
(i.e. assuming ${T_e} > {T_i}$, this condition can not be fulfilled for $s = -1$);
this is, in fact, not a realistic
physical situation \cite{PSbook}.
\bibitem{commentdiaw} See the discussions in Ref. \cite{AMS}, and definitions
therein, according to which
both $\alpha$, $\beta$ may be negative in the DIAW case.
Also note, for reference, that upon setting: $\alpha = -1/2$,
$\alpha' = 1/6$, $\beta = 1$ and $s = 1$ in (\ref{reducedeqs}), one
readily recovers the model equation system for the ion-acoustic waves, e.g.
exactly (1) in Ref. \cite{chin3} (also Ref. \cite{Shimizu} for $\sigma = 0$).
\bibitem{commentsignP}
Let $k_1$ the zero of $P$
i.e. the value of $k$ beyond which $P(k)$ changes sign:
$k_1 = 1 + \sqrt{4 + \frac{3 \beta}{\gamma \sigma}}$,
taking
finite values for $\sigma \ne 0$.
This does not contradict the remark made in Ref. \cite{chin3}  that
$P > 0$ for $k < 1$ (for $\theta = 0$ only).
\bibitem{commentlowQ}
This remark is in agreement with the ion-acoustic
wave case: see Eq. (41) in Ref. \cite{Shimizu}); as a matter of fact,
the factor $1/3$ therein is also exactly recovered here upon setting the
appropriate parameter values (see in Ref. \cite{commentdiaw}) into Eq. (\ref{lowQthetazero}).
\bibitem{commentFedele1} This result is immediately obtained from Ref.
\cite{Fedele}, by transforming the variables therein into our notation as
follows:
$x \rightarrow \zeta$, \, $s \rightarrow \tau$, \,
$\rho_m \rightarrow \rho_0$, \ $\alpha \rightarrow 2 P$, \
$q_0 \rightarrow - 2 P Q$, \ $\Delta \rightarrow L$, \
$E \rightarrow \Omega$, \ $V_0 \rightarrow u$.
\end{thebibliography}
\end{document}